\documentclass[conference]{IEEEtran}
\IEEEoverridecommandlockouts

\usepackage{pslatex}

\usepackage{cite}
\usepackage{hyperref}
\usepackage{amsmath,amssymb,amsfonts}
\usepackage{algorithmic}
\usepackage{graphicx}
\usepackage{textcomp}
\usepackage{xcolor}
\usepackage{todonotes}
\usepackage[utf8]{inputenc}	
\usepackage[T1]{fontenc}
\usepackage{textcomp}		
\usepackage{amsmath}		
\usepackage{fancybox}
\usepackage{anyfontsize}	
\usepackage{lipsum}
\usepackage{cleveref}
\usepackage{soul}
\usepackage{float}
\usepackage{booktabs}
\usepackage{listings}
\usepackage{tabularx}
\usepackage{subcaption}
\usepackage{makecell}
\usepackage{enumerate}
\usepackage[printonlyused]{acronym}
\def\BibTeX{{\rm B\kern-.05em{\sc i\kern-.025em b}\kern-.08em
    T\kern-.1667em\lower.7ex\hbox{E}\kern-.125emX}}
\usepackage[font={footnotesize}]{caption}

\begin{document}

\title{Scaling Analysis in a Multi-energy System\\

\thanks{This work has been supported by the ERIGrid 2.0 project of the H2020 Programme under Grant Agreement No. 870620.
\\
\copyright2023 IEEE.~https://doi.org/10.1109/EEE-AM58328.2023.10395068
Personal use of this material is permitted. Permission from IEEE must be obtained for all other uses, in any current or future media, including reprinting/republishing this material for advertising or promotional purposes, creating new collective works, for resale or redistribution to servers or lists, or reuse of any copyrighted component of this work in other works.}
}

\author{\IEEEauthorblockN{Jan Soeren Schwarz}
\IEEEauthorblockA{\textit{OFFIS} \\
Oldenburg\\
Germany\\
schwarz@offis.de}
\and
\IEEEauthorblockN{Minh Cong Pham}
\IEEEauthorblockA{\textit{CEA-Liten, INES} \\
Le Bourget du Lac\\
France\\
minh-cong.pham@cea.fr}
\and
\IEEEauthorblockN{Quoc Tuan Tran}
\IEEEauthorblockA{\textit{CEA-Liten, INES} \\
Le Bourget du Lac \\
France \\
quoctuan.tran@cea.fr}
\and
\IEEEauthorblockN{Kai Heussen}
\IEEEauthorblockA{\textit{Technical University of Denmark} \\
Lyngby \\
Denmark \\
kheu@dtu.dk}
}

\IEEEoverridecommandlockouts
\IEEEpubid{\makebox[\columnwidth]{979-8-3503-8106-1/23/\$31.00 ©2023 IEEE \hfill}
\hspace{\columnsep}\makebox[\columnwidth]{ }}
\maketitle
\IEEEpubidadjcol
\begin{abstract}
This paper presents a scaling study on the planning phase of a multi-energy system (MES), which is becoming increasingly prominent in the energy sector. The research aims to investigate the interactions and challenges associated with integrating heat and electrical systems and scaling their components. In this context, interaction between these two domains are investigated and the size of the distributed energy resources in the MES is scaled to examine the impact of sizing on the integrating networks and their controlling system. To achieve this, the paper uses sensitivity analysis and a meta-modeling technique, both incorporated in a toolbox for scaling analysis. These methodologies are validated through simulations, and the results obtained from the simulations can contribute to the advancement of MESs and their implementation in laboratory and field testing.
\end{abstract}


\section{Introduction}
\subsection{Context}

Efforts are being made worldwide, spanning multiple fronts from research to policy initiatives, to support the integration of renewable energy resources (RES) into the power system. One of the innovative concepts being pursued is the implementation of the Smart Grid. However, to successfully achieve ambitious environmental goals and ensure the availability of secure and affordable energy for present and future generations, it is crucial to develop comprehensive strategies that address all energy sectors, not just limited to electricity \cite{b1}. In addition, the planning of energy infrastructure expansions, including pipelines for natural gas or hydrogen, district heating networks and power transmission lines, should be approached from an integrated standpoint, considering their overall value across interconnected sectors \cite{b2}. Therefore, this necessitates the study of a multi-energy system (MES) that encompasses the interconnected relationships among various energy carriers and has the capacity to strategically design the composition of the future energy system.

By definition, MES refer to systems in which electricity, heat, cooling, fuels, transport, and other energy components interact with one another at different levels, such as within districts, cities, or regions \cite{b1}. Embracing MES presents a significant opportunity to enhance technical, economic, and environmental performance as compared to traditional energy systems, where sectors are treated separately or independently. This performance enhancement can occur both during the operational phase and in the planning stage. With the objective of cost optimization, authors in \cite{b15, b16, b17} propose sizing strategies for MES in the planning phase with energy management systems. There are several tools which can be useful for MES planning such as RETScreen \cite{b18}, DER-CAM \cite{b19} and eTransport \cite{b20}. However, the above mentioned tools and methods are not straightforward to identify
performance metrics that are capable to properly capture benefits that are relating to types of MES according to different criteria.

\subsection{Contribution}
In this paper, a MES is in the process of planning a local RES community, aiming to enhance the self-sufficiency of heat in the area through the use of a heat pump. However, considerations must also be given to address the limitations of the weak grid. To address this, a proposed solution involves implementing power-limiting voltage control, which restricts the consumption of the heat pump during times of low voltage at the end of the feeder or high demand on the grid. This case presents engineering challenges concerning the interactions across multiple domains, where resolving voltage and congestion issues will impact the operation of the heating plant, and vice versa. This paper focuses on conducting a quantitative analysis of two engineering concerns:
\begin{enumerate}
    \item Which interactions between the two domains (electricity and heat) hold importance and must be taken into consideration during the control design and sizing process?
    \item How much can the size of the tank contribute to a) alleviating voltage issues and b) maximizing the utilization of locally produced RESs in the area?
\end{enumerate}

The following two case studies delve into the two concerns mentioned in these scenarios. The first study explores the analysis of "domain expansion", i.e., effects which occur when considering the domains not only individually but integrated, while the second study focuses on the scaling analysis.

\subsection{Outline of the paper}
In this paper, the structure is as follows: Section 2 offers an introduction to the MES under investigation, encompassing the network configuration and control system. The sizing analysis methodology with a toolbox for the MES is illustrated in Section 3, along with the proposed utilization scenarios. Section 4 presents the simulation results and initiates a comprehensive discussion. Finally, in Section 5, the paper concludes by providing a summary and an outlook for future research in this area.

\section{Multi-energy system description}
\subsection{Multi-energy networks benchmark}
\begin{figure}[t]
    \centering
    \begin{subfigure}{.4\textwidth}
        \centering
        \includegraphics[width=\linewidth]{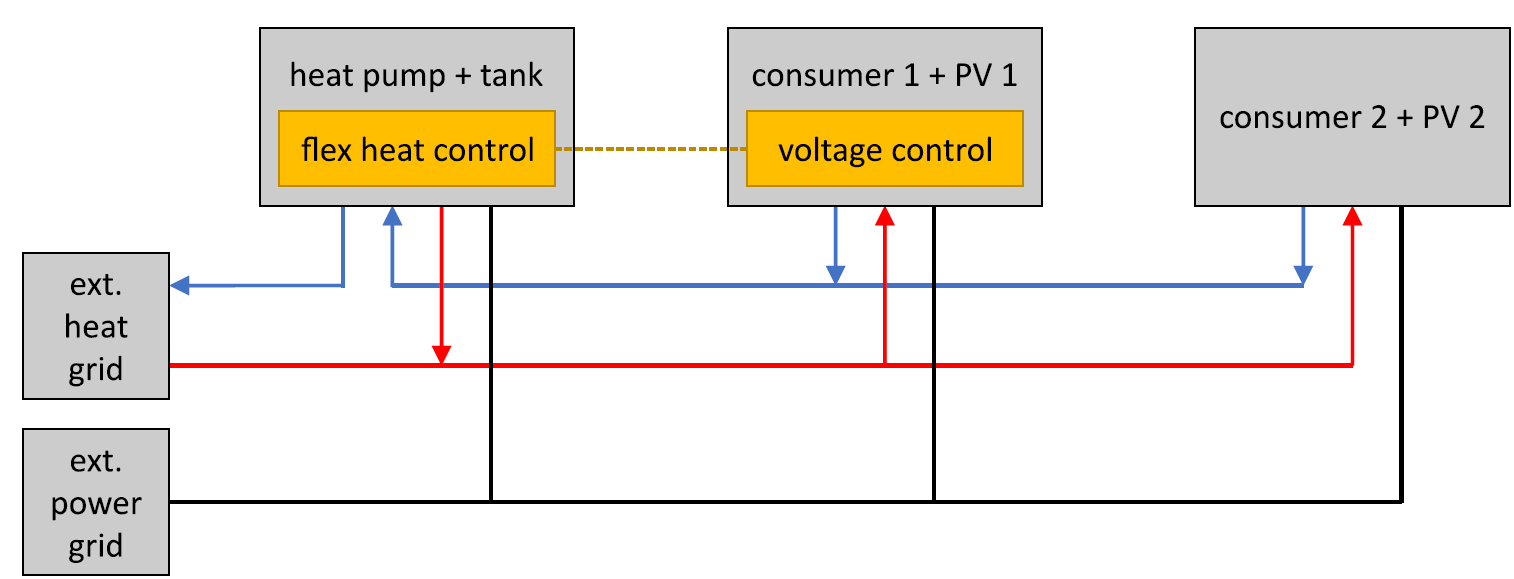}
        \caption{Overview of the overall MES.}
        \label{fig:me_benchmark_all}
    \end{subfigure}
     \begin{subfigure}{.4\textwidth}
        \centering
    \includegraphics[width=\linewidth]{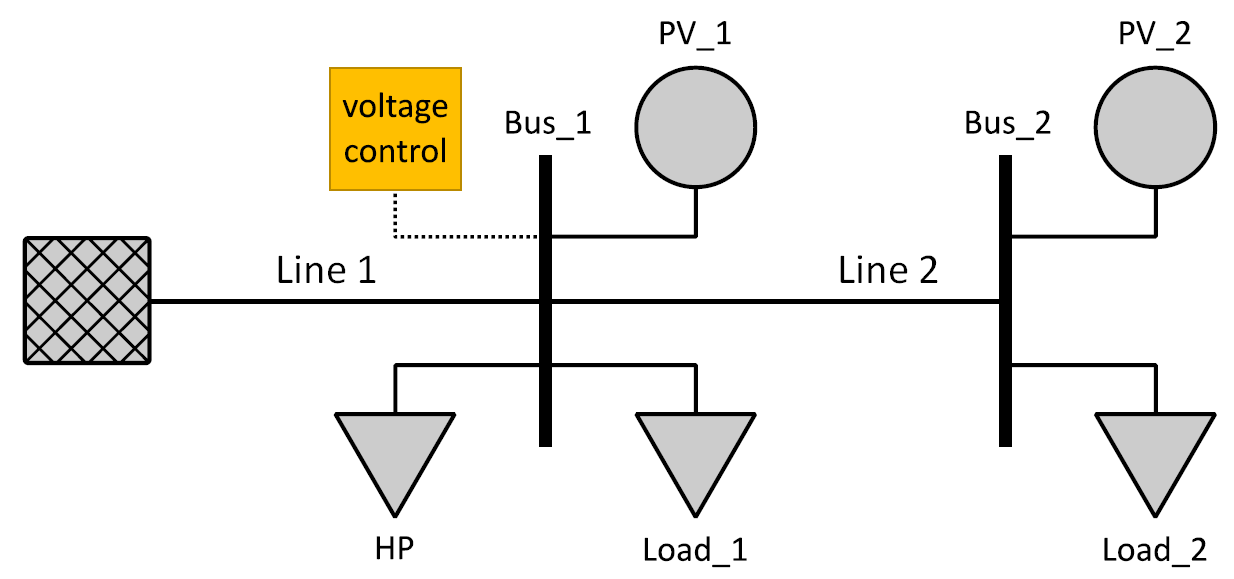}
        \caption{Detailed view of the components of the electrical sub-system.}
    \label{fig:me_benchmark_electrical}
    \end{subfigure}
    \begin{subfigure}{.42\textwidth}
        \centering
    \includegraphics[width=\linewidth]{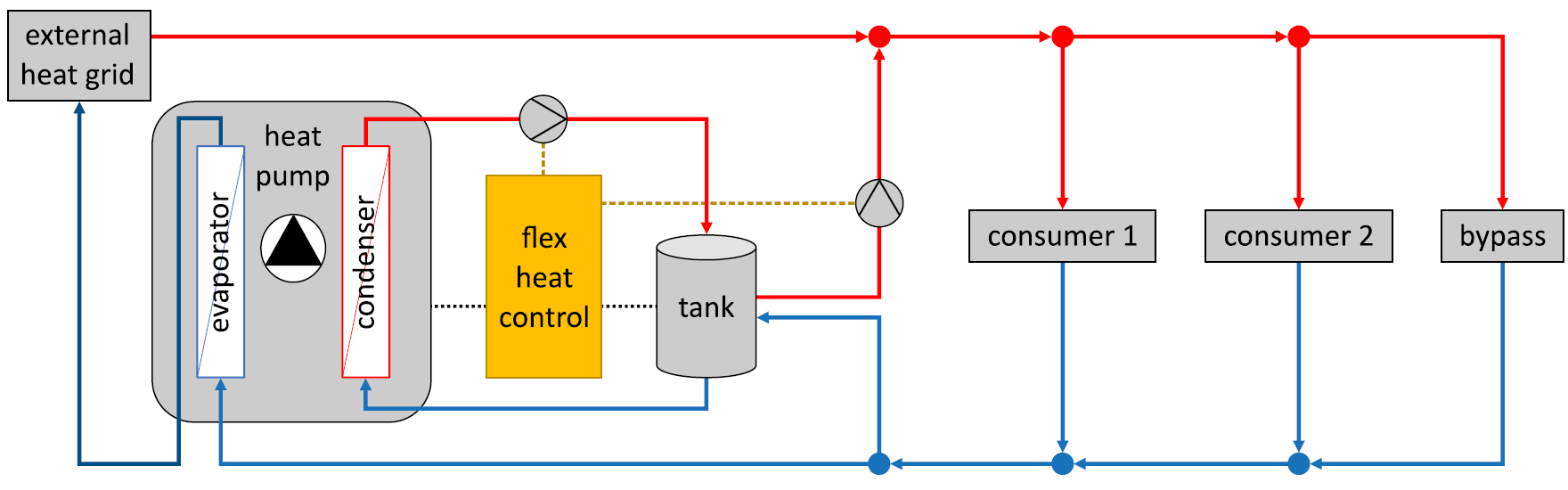}
        \caption{Detailed view of the components of the thermal sub-system.}
        \label{fig:me_benchmark_thermal}
    \end{subfigure}
    \caption{Multi-energy benchmark \protect\cite{b3}.}
    \label{fig:Multi-energy_benchmark}
\end{figure}
The numerical analyses conducted in this study will utilize a benchmark simulation model, which is described in Figure \ref{fig:Multi-energy_benchmark} and in references \cite{b4,b5}. Its code is available at \cite{b3}. The case under consideration involves a co-simulation scenario that incorporates both electricity and district heating infrastructure. This benchmark outlines a standard configuration for a MES coupling application. It involves a power-to-heat facility that acts as a connection point between a low-voltage distribution network and a localized section of a heating network. The power-to-heat facility serves a dual purpose: utilizing surplus PV generation from the local area and enhancing the stability of the electrical network, while simultaneously contributing to the thermal network's supply. The benchmark system utilized in this study simulates a suburban setting that incorporates a substantial number of PV installations. The chosen scenario involves the implementation of a Local Energy Community (LEC) with the aim of harnessing local excess PV generation to operate a power-to-heat facility. The system configuration comprises several sub-systems and components, including \cite{b4}:
\begin{itemize}
    \item Electrical LV distribution network: Two transmission lines (0.3 km each), connected to an external power grid.
    \item Thermal network: Three pipes (0.5 km each), connected to an external district heating grid.
    \item Consumption: Two consumers, each representing the aggregated loads (electrical and thermal) of a residential neighbourhood and connected to both networks.
    \item Generation: Two PV systems (one of 150 $kW_{el,peak}$ and one of 50 $kW_{el,peak}$).
    \item Power-to-heat facility: heat pump (max. 100 $kW_{el}$) connected to a thermal tank (100 $m^{3}$) feeding into the thermal network.
\end{itemize}
\subsection{Multi-energy control system}
\begin{figure}[t]
    \centering
    \begin{subfigure}{.42\textwidth}
        \centering
        \includegraphics[width=\linewidth]{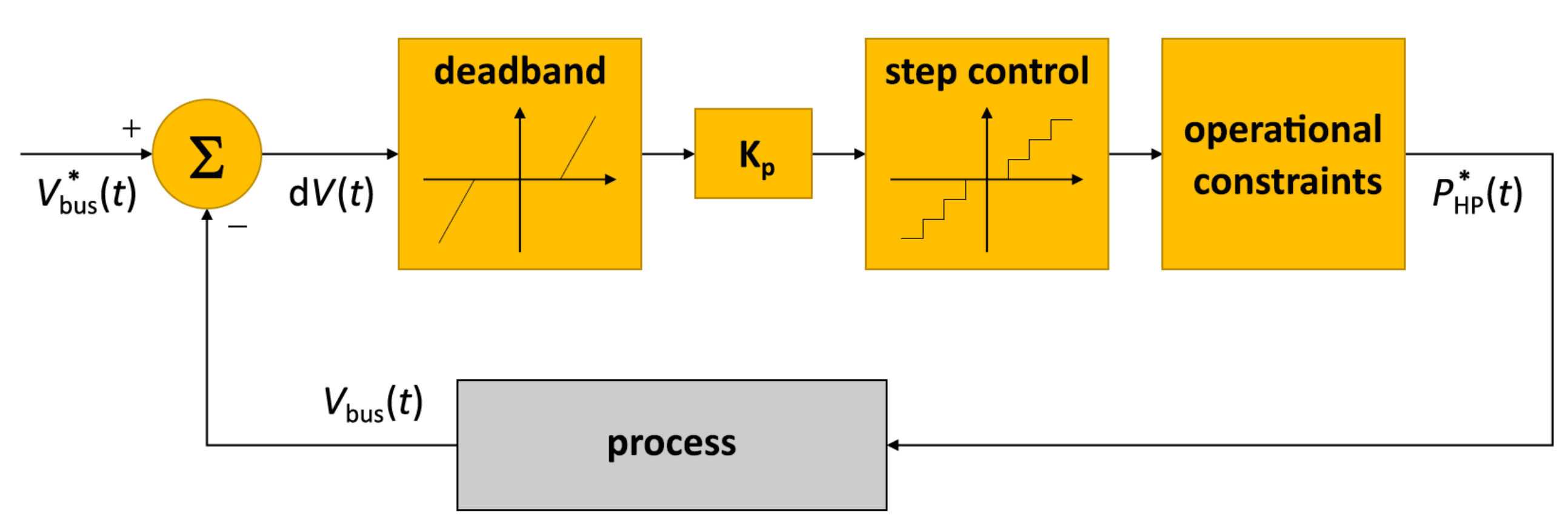}
        \caption{Schematic view of the voltage control algorithm.}
        \label{fig:voltage control}
    \end{subfigure}
    \begin{subfigure}{.42\textwidth}
        \centering
    \includegraphics[width=\linewidth]{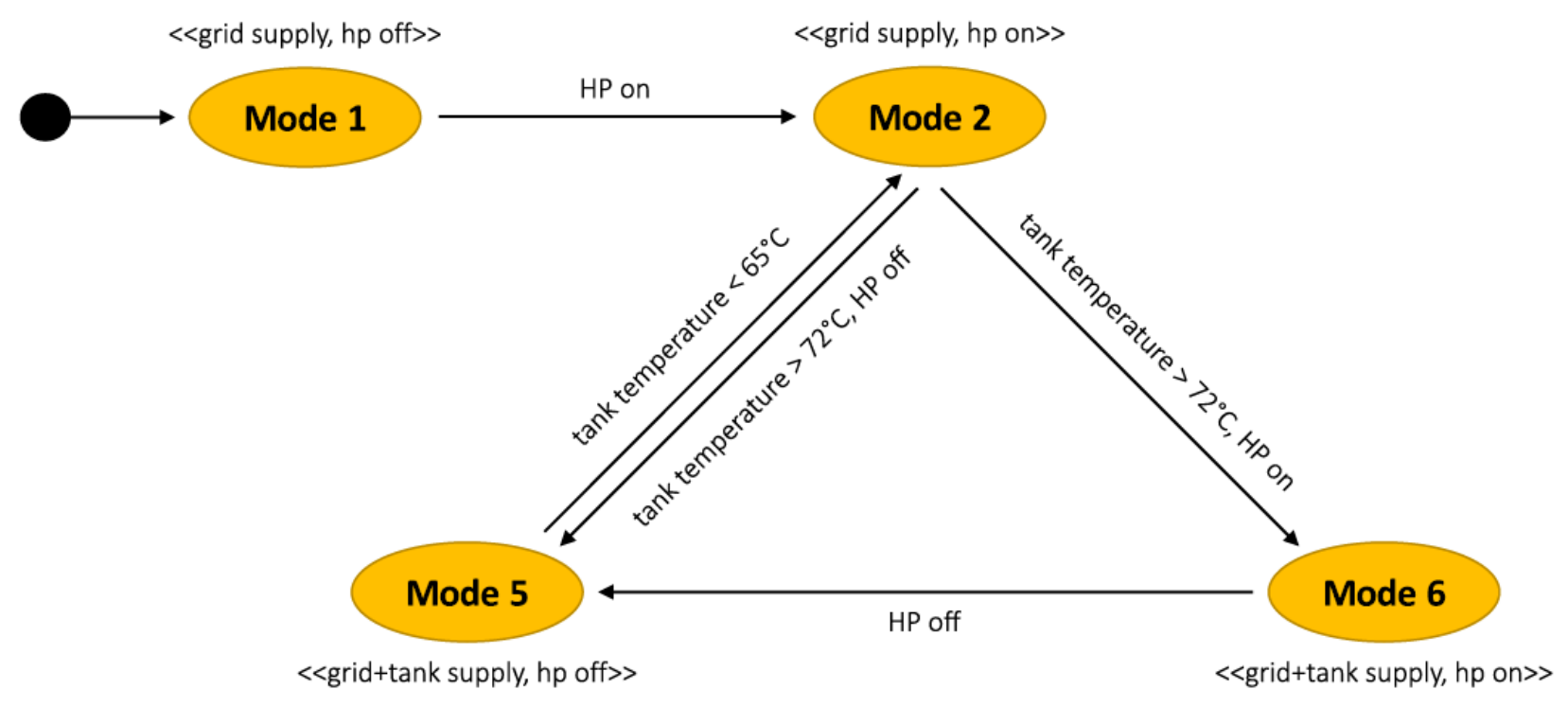}
        \caption{Schematic view of the flex heat control algorithm.}
    \label{fig:heat control}
    \end{subfigure}
    \caption{Control algorithm for the multi-energy benchmark \protect\cite{b4}.}
    \label{fig:Control_Multi-energy_benchmark}
\end{figure}

The benchmark's case study tackles concerns surrounding self-consumption within a LEC. To achieve autonomy, the PV system must be appropriately sized. However, if there is a disparity between energy demand and supply from the PV systems, it can result in a substantial increase in voltage within certain sections of the power grid. Consequently, synchronization between consumption and generation becomes essential to guarantee power quality and prevent disruptions caused by exceeding voltage limits. To achieve this objective, a straightforward voltage control method is implemented. The voltage at Bus 1 in Figure \ref{fig:me_benchmark_electrical} is continuously monitored, and the power consumption target of the heat pump is adjusted accordingly to maintain the voltage within acceptable boundaries. Figure \ref{fig:voltage control} illustrates the voltage control algorithm associated with this approach. On the other hand, the thermal sub-system employs a dedicated controller scheme known as flex heat control. This controller determines whether the heat supply should be entirely provided by the external grid or if the power-to-heat facility should discharge the tank to support it. The flex heat controller operates in various modes based on measurements of the storage tank temperature and the heat pump's power consumption threshold. These modes in Figure \ref{fig:heat control} are presented as a state machine, where each state corresponds to a specific operational mode.

\section{Methodologies for analysis}

\subsection{Toolbox for Analysis}
\begin{figure}[t]
    \centering
    \includegraphics[width=\linewidth]{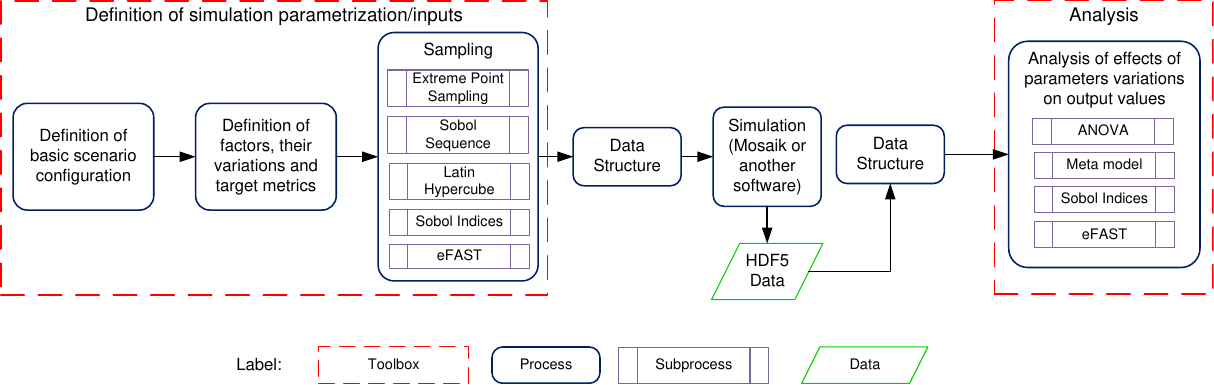}
        \caption{Flowchart of the proposed toolbox for the scaling sensitivity analysis \cite{jra_1_2}.}
    \label{fig:Toolbox Process in Summer School}
\end{figure}

In order to analyse the sensitivity of the system scaling in the MES, the ERIGrid 2.0 project-based toolbox is utilized to assess the sensitivity of system scaling in the MES. The toolbox was initially based on an a Design of Experiments (DoE) tutorial example scenario \cite{b7} and additional descriptions of the integration of DoE into the Holistic Testing Description (HTD) \cite{doe_htd} from the ERIGrid 1.0 project. The initial tutorial was re-engineered and the SALib toolbox \cite{b9} was integrated, which provides various sampling and analysis methods, which added more sensitivity analysis sampling strategies. The toolbox comprises three primary components, described in Figure \ref{fig:Toolbox Process in Summer School}, which provides a visual representation of the toolbox structure and its corresponding procedures. A more detailed description can be found in \cite{jra_1_2} and the code is available at GitHub\footnote{\href{https://github.com/ERIGrid2/toolbox\_doe\_sa}{https://github.com/ERIGrid2/toolbox\_doe\_sa}}.

The first part consists of a toolbox for defining simulation parameters and inputs. It involves defining the basic scenario configuration and specifying factors, their variations, and target metrics. These variations are used to generate parameter combinations, which are also called recipes, based on different sampling approaches. Next, the mosaik co-simulation framework \cite{b8} is utilized for simulation, directly employing the predefined recipes for multiple simulation runs. However, the toolbox is not limited to specific software and can be used with other simulation tools or hardware laboratory experiments by providing the recipes as JavaScript Object Notation (JSON) data. Once the simulation framework receives the parameterizations generated in the previous step, it incorporates the variable values into the model to calculate the system response. The simulation results are then used as input for analysis module of the toolbox. The toolbox employs the adequate statistical methods for the used sampling approach to examine the effects of input parameter variations on the response of the modeled system. Moreover, the toolbox offers configurable plots that visually represent the relationship between input and output variations. These plots can easily be customized by the user.
\subsection{Interactions between the two domains}
To identify the significant inter-domain interactions, the initial technique employed is the One (factor)-at-a-Time (OAT) sensitivity analysis \cite{oat}. It would also be feasible to begin directly with a Global Sensitivity Analysis (GSA) method, like sobol indices, which would involve simulating combinations of all factor variations. But the OAT approach, as the name suggests, only alters one factor at a time and requires fewer simulation runs. With OAT approach, each chosen factor's impact is evaluated individually by sequentially changing its values between consecutive assessments, starting from an initial base value of the input vector. Compared to a full factorial design that evaluates all possible combinations of input factor values, the OAT design requires significantly fewer model or system evaluations. This preliminary analysis proves advantageous as it saves time and effort by identifying non-influential factors and disregarding their impact on response variability in subsequent analyses.

The sobol indices approach for GSA is based on the variance of the simulation results over varied inputs\cite{sobol,saltelli}.
The direct effect of only varying one factor alone is represented by the first-order sensitivity coefficient (S1), the higher order effects from interactions with other factors are represented by the total effect index (ST).

\begin{table}[t]
    	\centering
    	\caption{Target Metrics for Simulation Cases Study}
    	\label{tab:target_metrics_for_Case_5}
    	\renewcommand{\arraystretch}{1.7} 
        \begin{tabular}{p{5 cm}<{\centering}|p{2 cm}<{\centering}}        
    		\hline
    		\hline      
    		\textbf{Target Metric} & \textbf{Unit}\\
    		\hline            
            Maximum voltage bus 2 &  p.u. \\[1.2pt] 
            Maximum line loading line 0 &  \% \\[1.2pt]  
            Heat pump average COP & Unit \\[1.2pt] 
            Self-consumption electricity &  MWh \\[1.2pt] 
            Minimum supply temperature (Critical node temperature) &  $^\circ\text{C}$  \\[1.2pt] 
            Imported heat &  GWh \\[1.2pt]        
    		\hline      
    	\end{tabular}
\end{table}
\subsection{Tank size scaling problem analysis}
In this part, the objective is not to perform sensitivity screening for domain extension but rather to demonstrate a potential scaling analysis through a meta-modeling (surrogate models) approach, which involves constructing a meta-model using the simulation results to depict the influence of variations on the outcomes. A surrogate model refers to a simple mathematical function that approximates the underlying complex process function, denoted as y = f(x), using input-output samples. This particular approach can be further expanded to include optimization, as demonstrated in the work of \cite{b6}. One MES's component that is focused for investigating this scaling analysis is the Hot Water Tank (HWT). The tank volume can be adjusted based on the parameters of height and inner diameter. However, in the subsequent analysis, only the inner diameter is utilized, while the height remains fixed.

\section{Simulation results and discussion}

\subsection{Analysis of Inter-domain Causality and Sensitivity}\label{AA}
\begin{figure}[t]
    \centering
        \includegraphics[width=\linewidth]{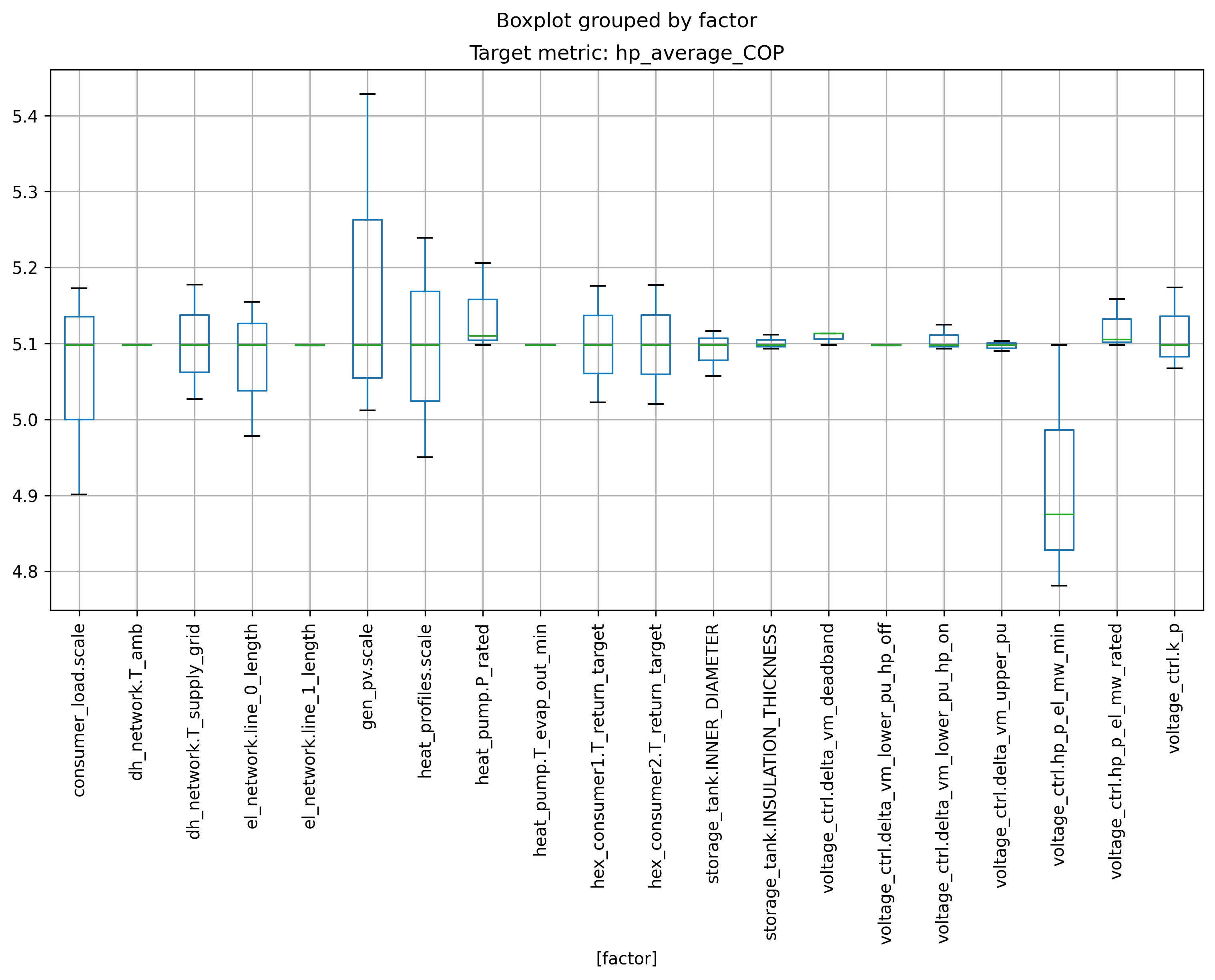}
        \caption{One (factor)-at-a-Time (OAT) analysis for target metric average Coefficient of Performance (COP) of the heat pump.}
    \label{fig:OAT_COP}
\end{figure}

In this section, a simulation is described using the standard configuration as a baseline, and then each factor is individually varied within its defined range, from minimum to maximum. Based on the objectives mention in the Contribution section, a list of target metrics is proposed and presented in the Table \ref{tab:target_metrics_for_Case_5}. The impact of these variations on the target metrics is analyzed based on the results obtained from these three simulation runs per factor. The impact of the factors on the target metric, specifically the average Coefficient of Performance (COP) of the heat pump, is exemplified in Figure \ref{fig:OAT_COP}.

\begin{figure}[b]
    \centering
        \includegraphics[width=\linewidth]{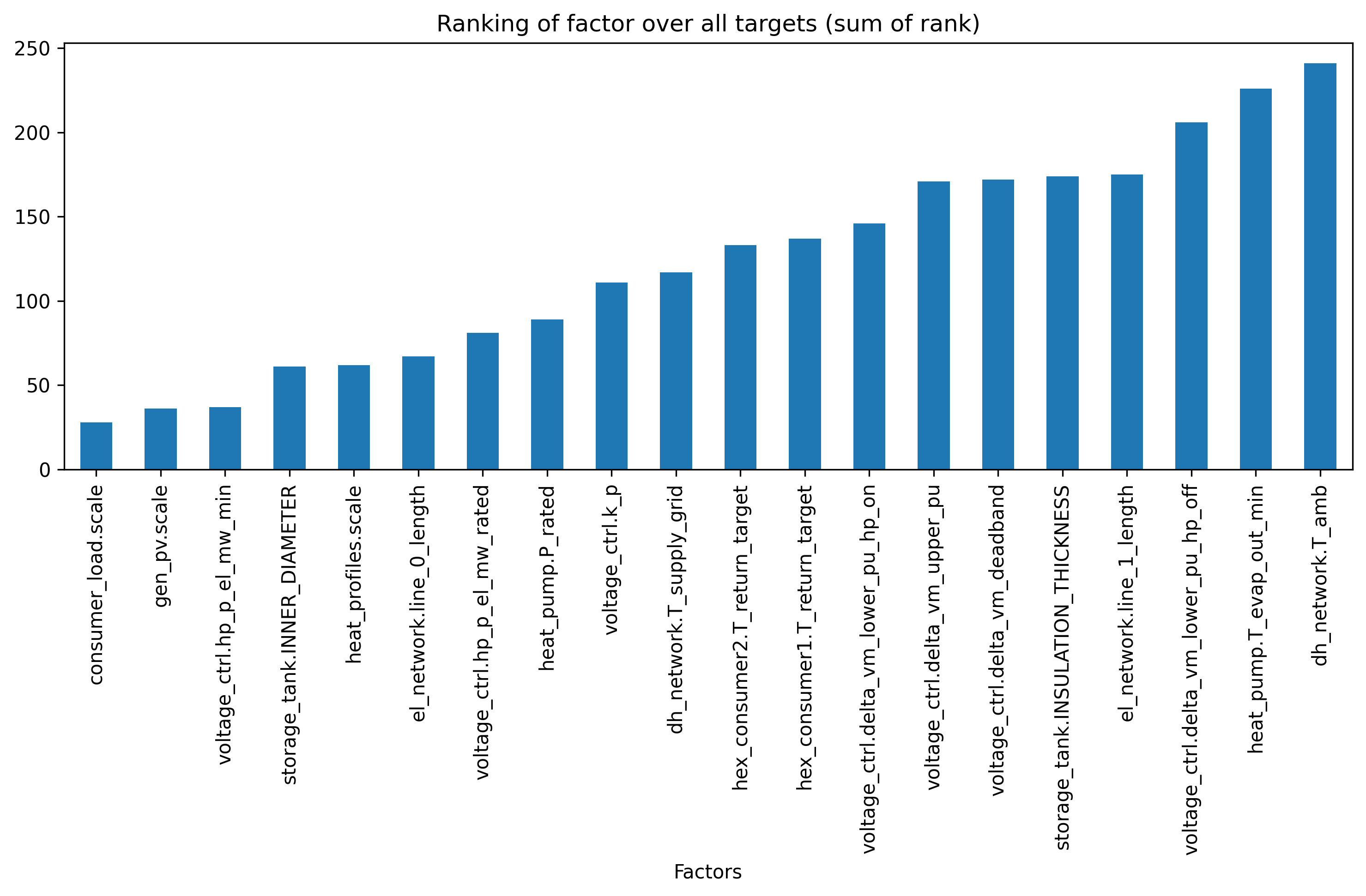}
        \caption{Ranking of the impact of factors over all target metrics.}
    \label{fig:Ranking of the impact of factors over all target metrics}
\end{figure}

Figure \ref{fig:Ranking of the impact of factors over all target metrics} depicts the ranking calculation for comprehensive understanding of a single factor's overall impact. Rankings result from sorting factors for each target metric, using variance across three simulation runs. This analysis, albeit small-sample and focusing on direct effects, aids users in discerning factors for deeper scrutiny or with minimal impact. Notably, domain expertise is vital due to results' sensitivity to parameter range adjustments, potentially necessitating an iterative approach to establish pertinent sensitivity analysis ranges.
To conduct further investigation, there are two primary areas of focus: (1) examination of design parameters that are of interest from an engineering perspective, and (2) exploration of scenario parameters that may be crucial for up-scaling. Within the scenario parameters, specific attention is given to the heat profile, PV scaling, and power of the heat pump due to their high impact indicated by the OAT sensitivity analysis. While the consumer load scaling also demonstrated a significant impact, it was not further analyzed as its effect is technically similar to PV scaling. Among the design parameters, the inner diameter of the HWT, the minimum operating point of the heat pump, and the proportional term $K_{p}$ of the voltage controller are selected for further investigation.

\begin{figure}[b]
    \centering
    \begin{subfigure}{.24\textwidth}
        \centering
        \includegraphics[width=1\textwidth]{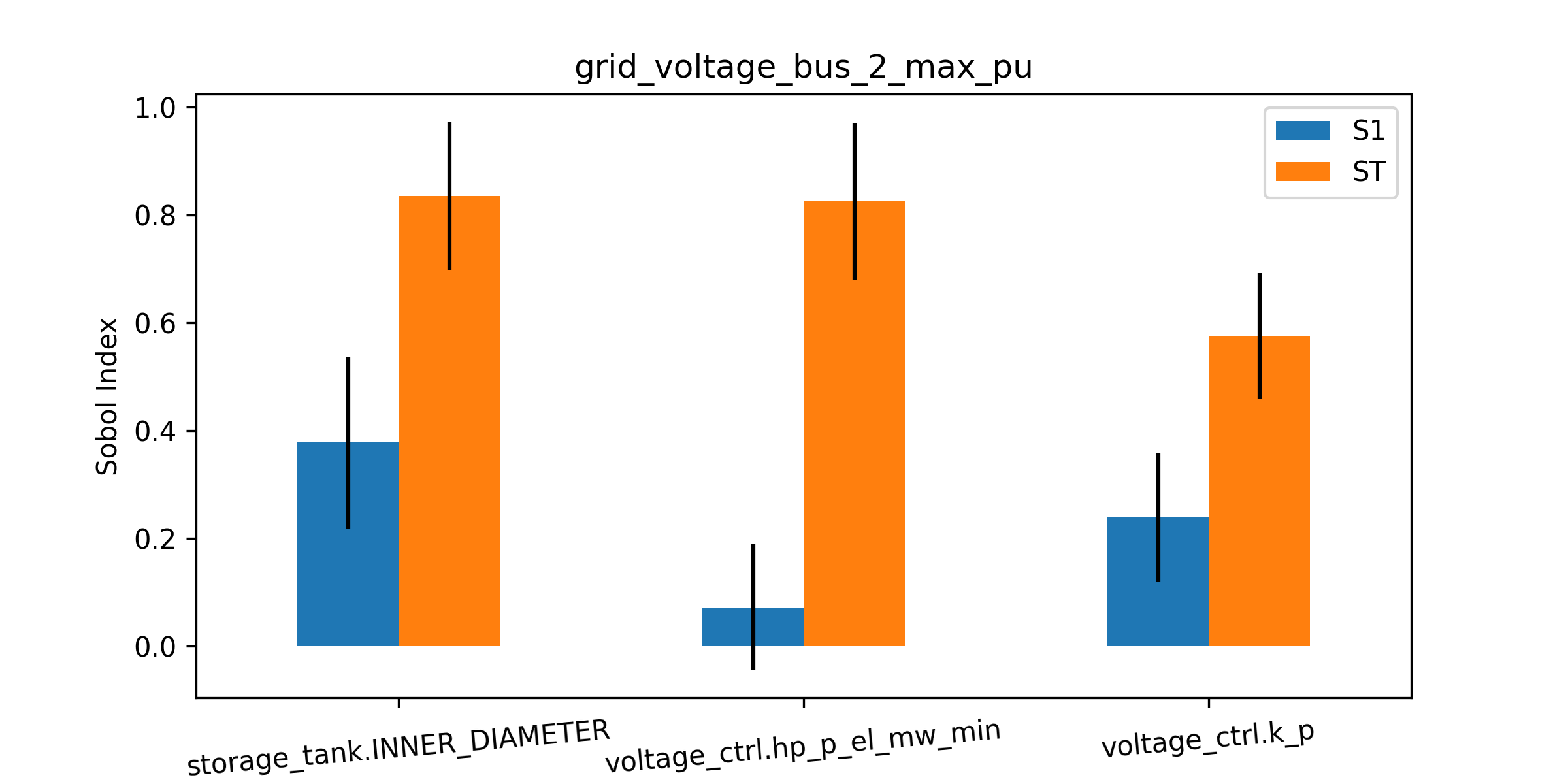}
        \caption{Sobol indices for maximum voltage at bus 2 of electricity.}
        \label{fig:Sobol indices for maximum voltage at bus 2 of electricity}
    \end{subfigure}
    \begin{subfigure}{.24\textwidth}
        \centering
        \includegraphics[width=1\textwidth]{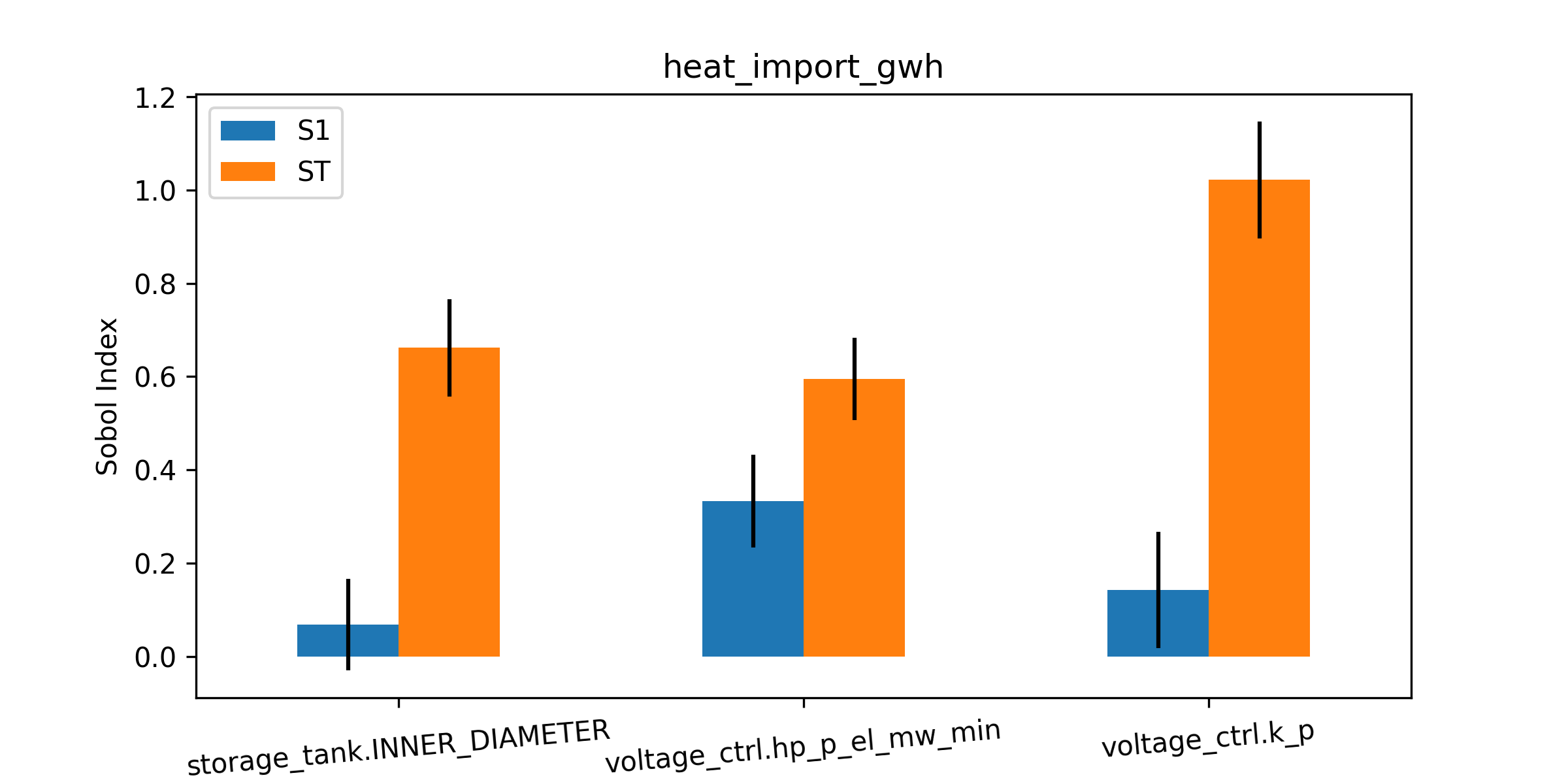}
        \caption{Sobol indices for heat import from external grid.}
        \label{fig:Sobol indices for heat import from external grid}
    \end{subfigure}
    \newline
    \begin{subfigure}{.24\textwidth}
        \centering
        \includegraphics[width=1\textwidth]{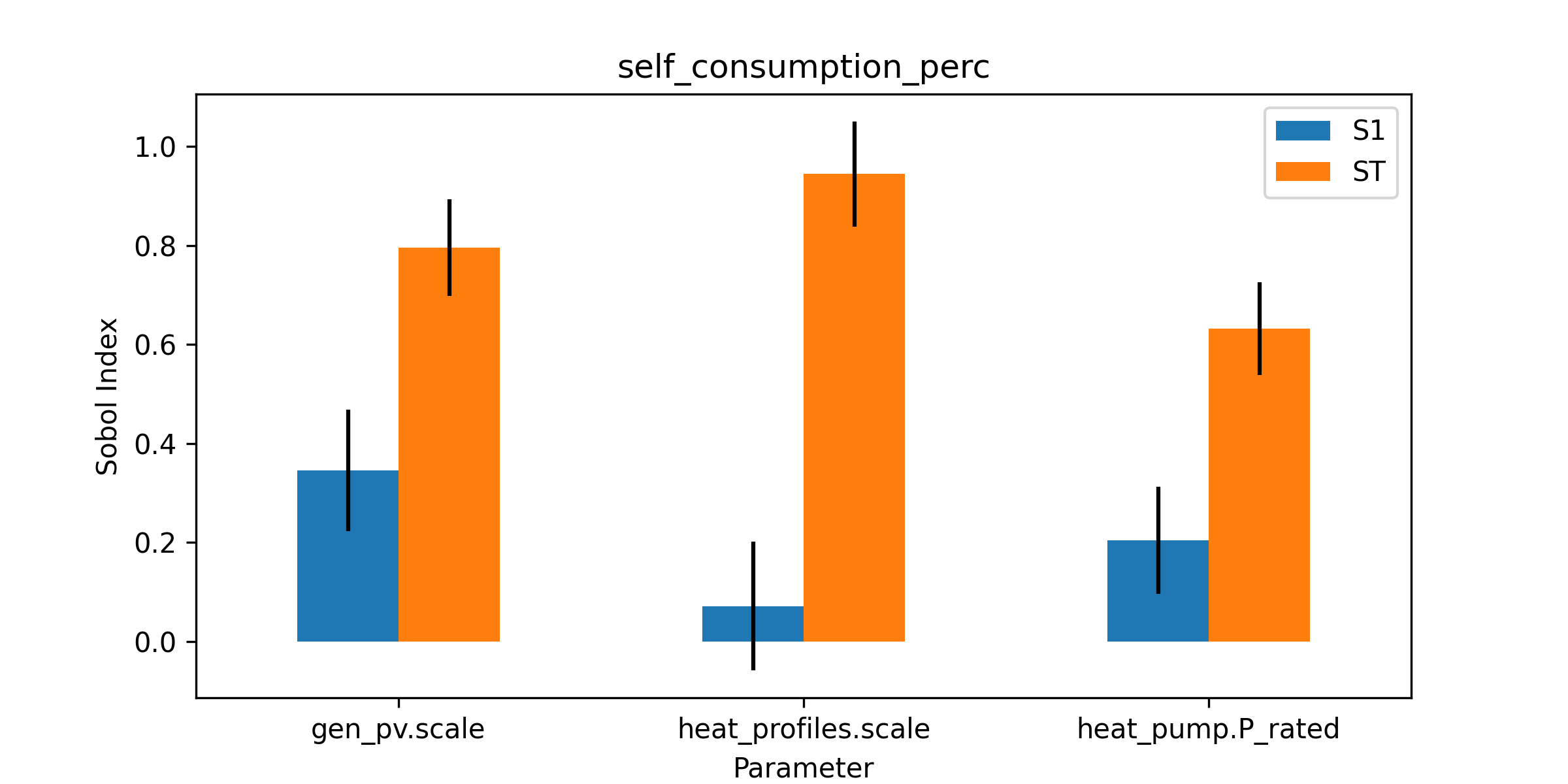}
        \caption{Sobol indices for self-consumption of electricity.}
        \label{fig:Sobol indices for self-consumption of electricity}
    \end{subfigure}
    \begin{subfigure}{.24\textwidth}
        \centering
        \includegraphics[width=1\textwidth]{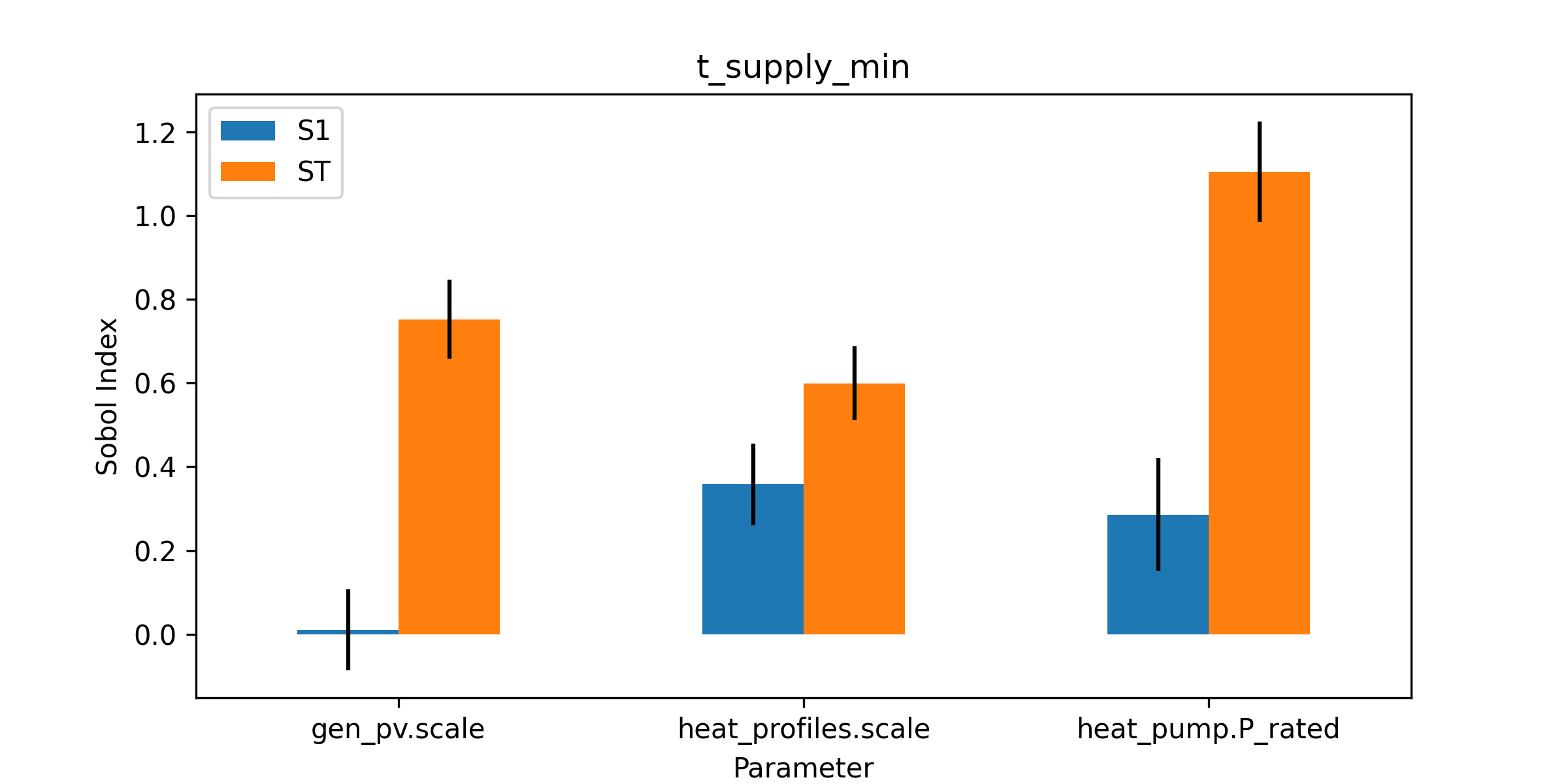}
        \caption{Sobol indices for minimum supply temperature.}
        \label{fig:Sobol indices for critical node temperature of heat}
    \end{subfigure}
    \caption{Sobol indices for different target metrics resulting from sampling of design (a-b) and scenario (c-d) parameters.}
    \label{fig:Sobol_all}
\end{figure}

After that, a GSA was conducted to understand deeper the impact of chosen factors on target metrics. Sobol indices sampling method was employed for 1024 samples. The resulting simulation data was analyzed using Sobol indices to calculate the first-order sensitivity coefficient (S1) and total effect index (ST). The analysis results are presented in Figure \ref{fig:Sobol_all}. 

The Sobol indices provide insights into the impact of factor variations on the target metrics, allowing for the ordering of factors based on their effects. Furthermore, the confidence of the results is calculated with a 95 $\%$ confidence interval, represented by the black lines. In Figure \ref{fig:Sobol indices for maximum voltage at bus 2 of electricity} and \ref{fig:Sobol indices for heat import from external grid}, the Sobol indices were calculated for the three selected design parameters. Focusing on the first-order effects (blue bars), it can be observed that the diameter of the HWT and the proportional term $K_{p}$  of the voltage controller have the greatest impact on the bus voltage (Figure \ref{fig:Sobol indices for maximum voltage at bus 2 of electricity}). On the other hand, the minimum operating point of the heat pump predominantly influences the heat import to the external grid (Figure \ref{fig:Sobol indices for heat import from external grid}).

In Figure \ref{fig:Sobol indices for self-consumption of electricity} and \ref{fig:Sobol indices for critical node temperature of heat}, the Sobol indices were computed for the three chosen scenario parameters. The results indicate that the scaling of PV has the most substantial effect on electricity self-consumption, followed by the power of the heat pump, while the impact of heat profile scaling is relatively minor. Conversely, when considering the critical node temperature, the influence of PV scaling is nearly negligible, while the other two factors exhibit more prominent effects.

\subsection{Tank size scaling problem analysis}
In this section, the same multi-energy benchmark simulation scenario serves as the base. The focus here is on illustrating a potential scaling analysis using a meta-modeling approach.

The first analysis considers the inner diameter of the HWT, while the height remains fixed at 7.9m. The diameter is varied between 1 and 8 meters, corresponding to a volume range of approximately 620 to 40000 liters. With a temperature difference of 10 Kelvin and a 100kW heat pump rated capacity, this volume range provides storage capacities from 1 minute to 9 hours of full thermal load capacity.

\begin{figure}[t]
    \centering
    \begin{subfigure}{.24\textwidth}
        \centering
        \includegraphics[width=1\textwidth]{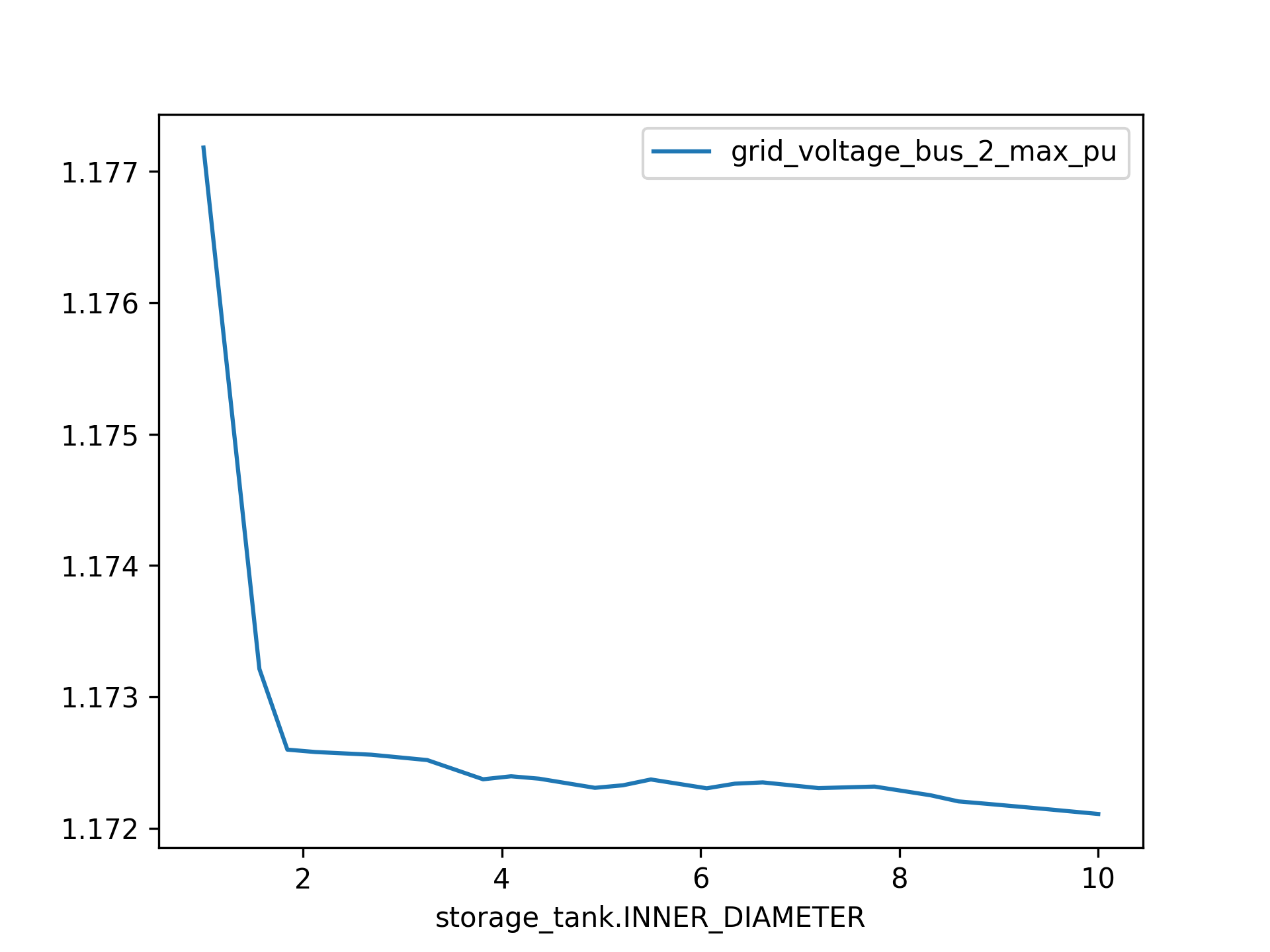}
        \caption{Maximum voltage at bus 2.}
        \label{fig:HWT_Maximum voltage at bus 2}
    \end{subfigure}
    \begin{subfigure}{.24\textwidth}
        \centering
        \includegraphics[width=1\textwidth]{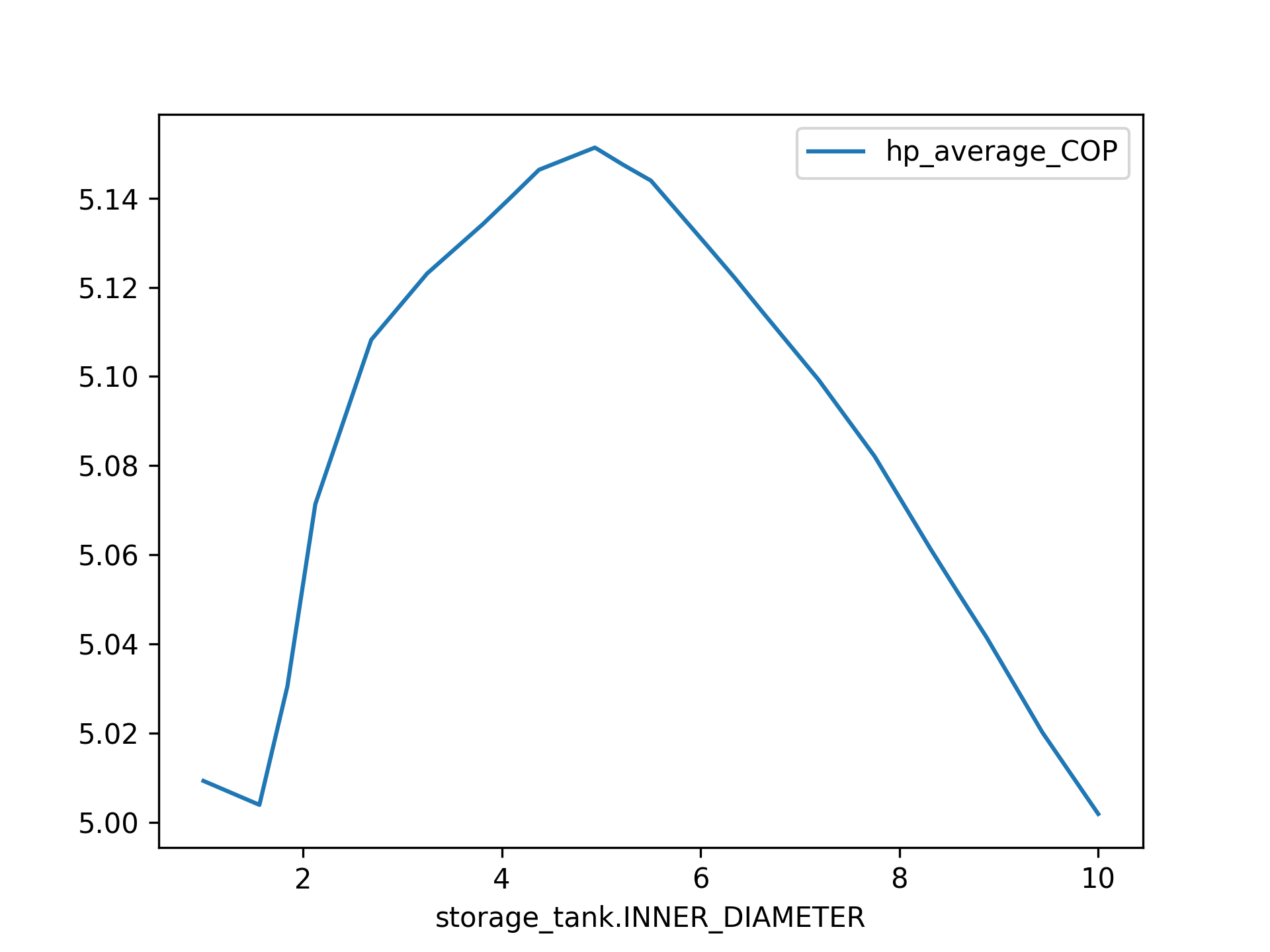}
        \caption{Average COP of heat pump.}
    \label{fig:HWT_Average COP of heat pump}
    \end{subfigure}
    \newline
    \begin{subfigure}{.24\textwidth}
        \centering
        \includegraphics[width=1\textwidth]{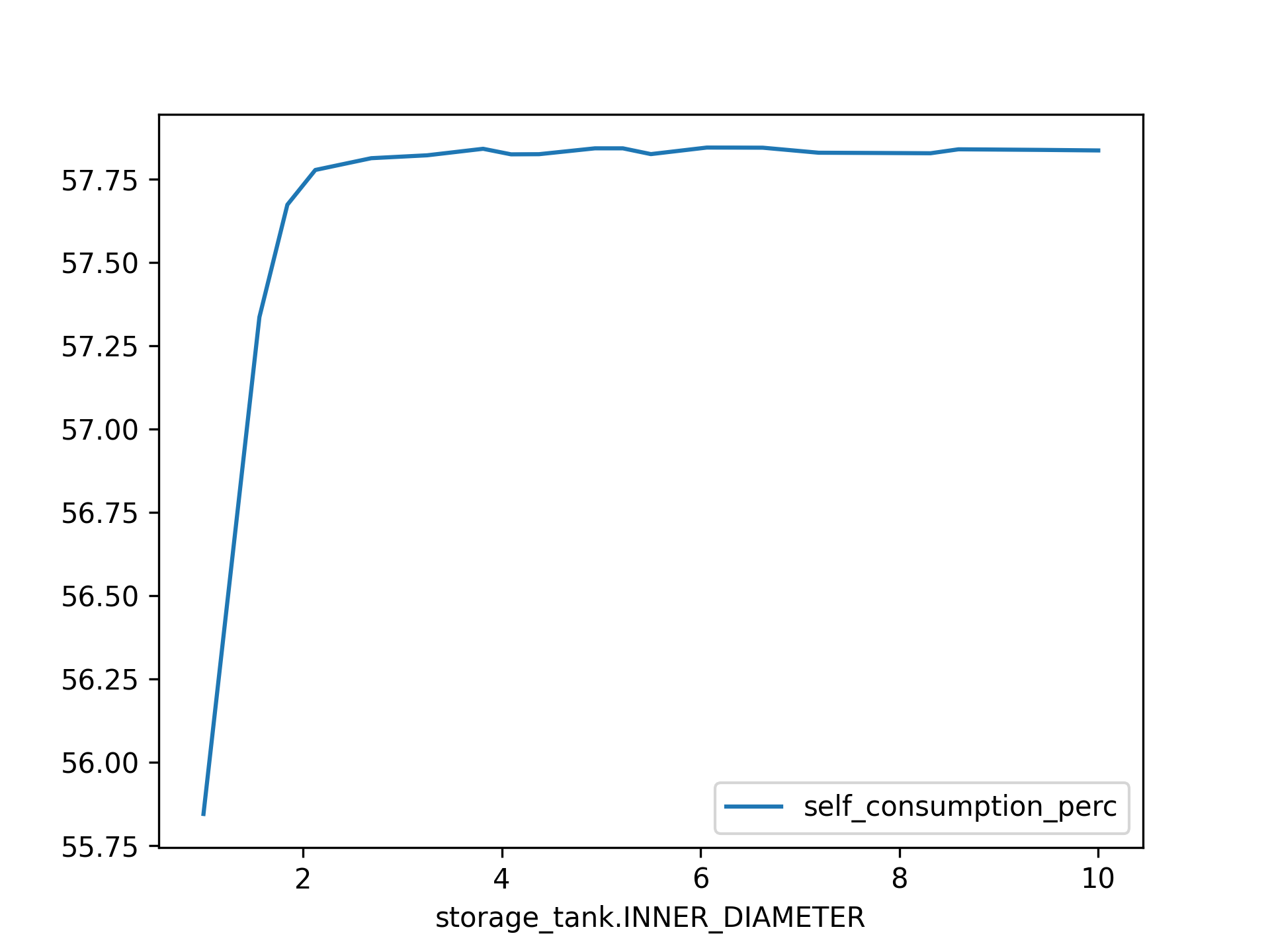}
        \caption{Self-consumption in percent.}
        \label{fig:HWT_Self-consumption in percent}
    \end{subfigure}
    \begin{subfigure}{.24\textwidth}
        \centering
        \includegraphics[width=1\textwidth]{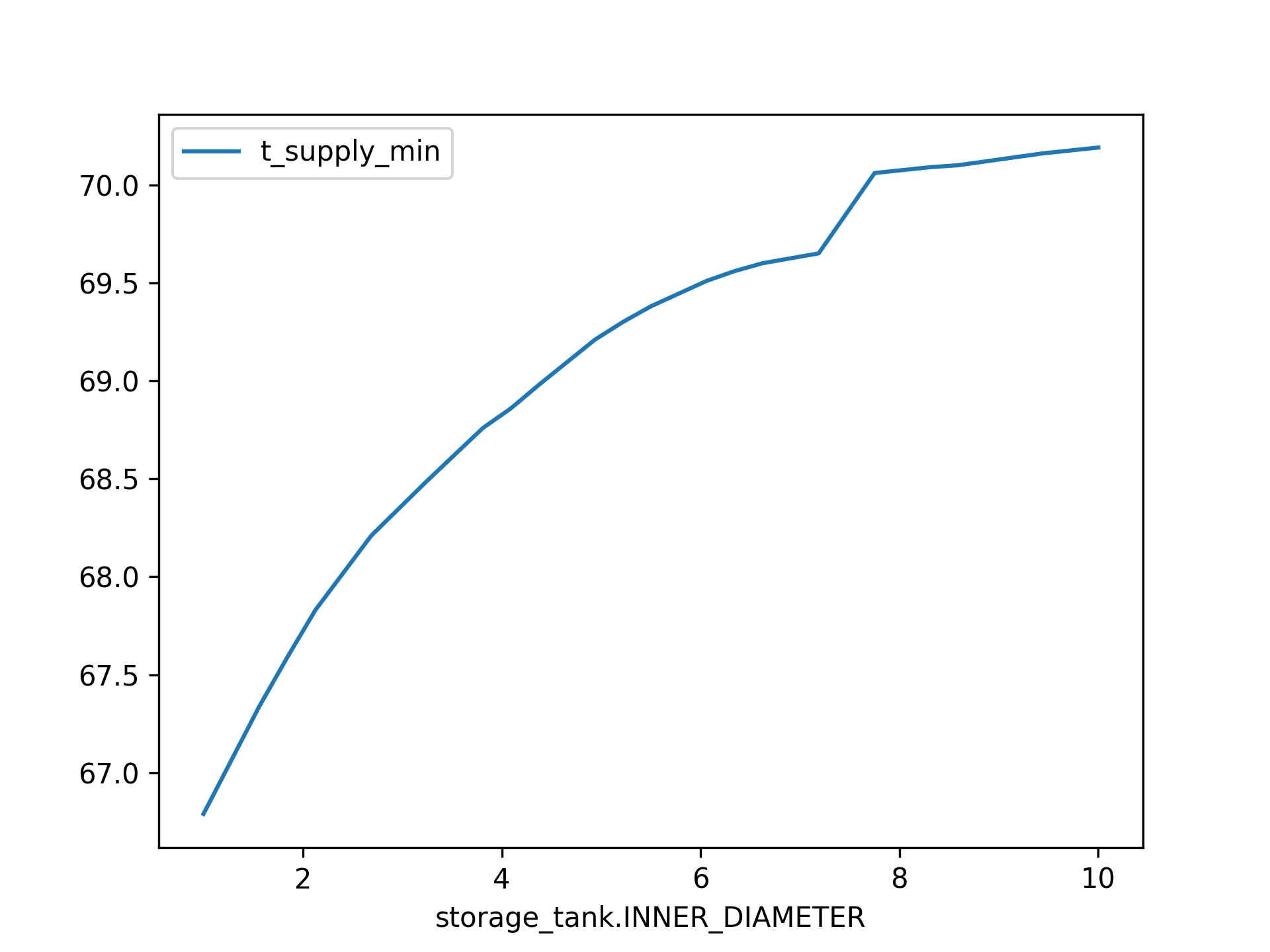}
        \caption{Minimum supply temperature.}
    \label{fig:HWT_Minimum supply temperature (critical node temperature)}
    \end{subfigure}
    \caption{Impact of the inner diameter scaling of the Hot Water Tank (HWT) to four target metrics.}
    \label{fig:Impact of variations in the inner diameter of the HWT to four target metrics.}
\end{figure}

Figure \ref{fig:HWT_Maximum voltage at bus 2} shows that for voltage, as well as Figure \ref{fig:HWT_Self-consumption in percent} for self-consumption, the inner diameter of the HWT has a significant impact, particularly for diameters smaller than 2 meters. Enlarging the HWT still yields a slightly positive impact for larger sizes. In Figure \ref{fig:HWT_Average COP of heat pump}, the average COP of the heat pump demonstrates an optimum around a 5-meter inner diameter. On the other hand, the critical node temperature benefits from an increasing diameter, which saturates at approximately 10 meters (Figure \ref{fig:HWT_Minimum supply temperature (critical node temperature)}). These findings may provide initial insights for scaling the HWT size. However, it's important to note that other scaling effects can also influence the scaling of the HWT size. Therefore, further analysis will explore the scaling of the HWT in conjunction with other scaling factors.

\begin{figure}[htbp]
    \centering
    \begin{subfigure}{.24\textwidth}
        \centering
        \includegraphics[width=1\textwidth]{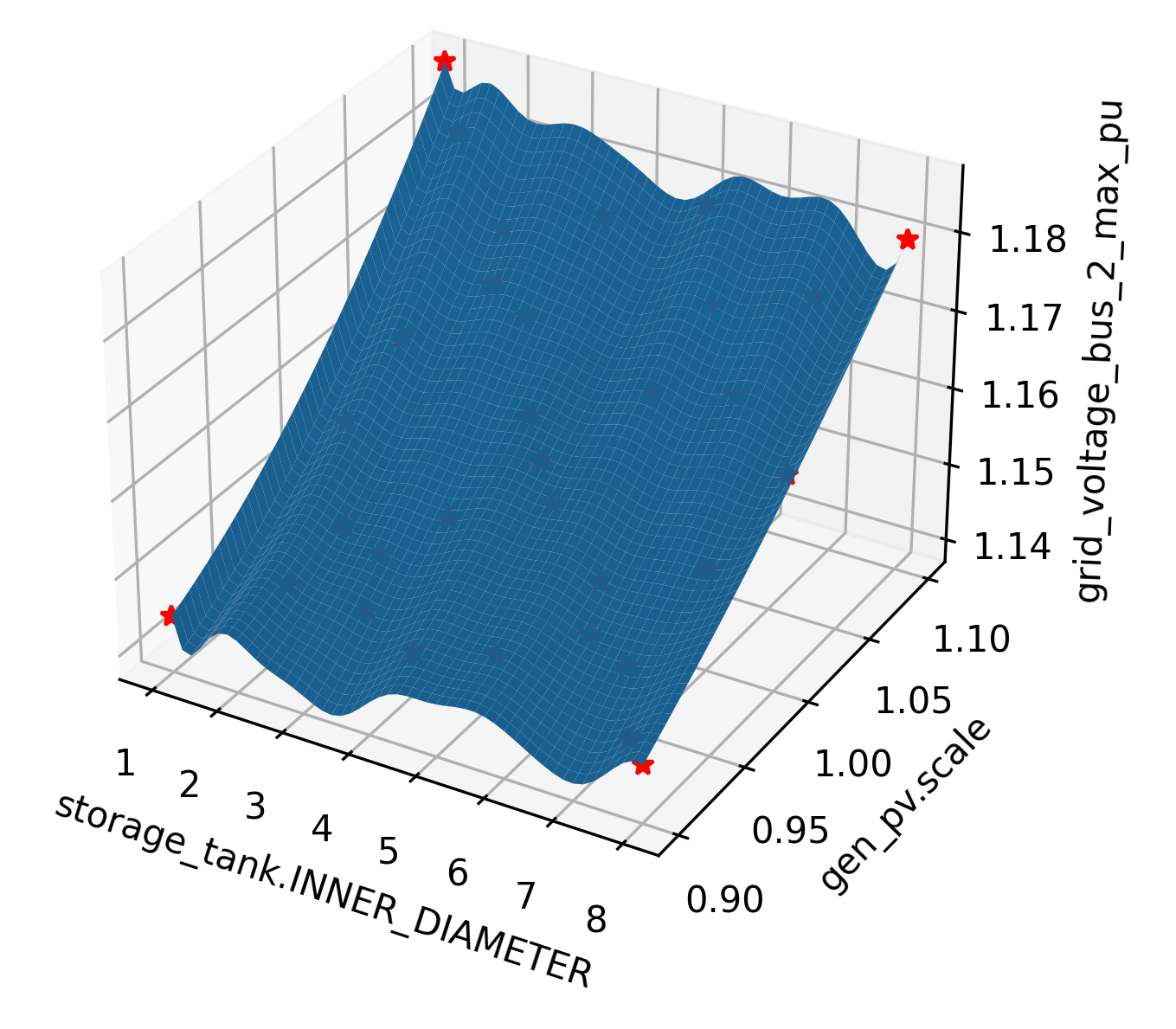}
        \caption{Maximum voltage at bus 2.}
        \label{fig:meb_scaling_pv_voltage}
    \end{subfigure}
    \begin{subfigure}{.24\textwidth}
        \centering
        \includegraphics[width=1\textwidth]{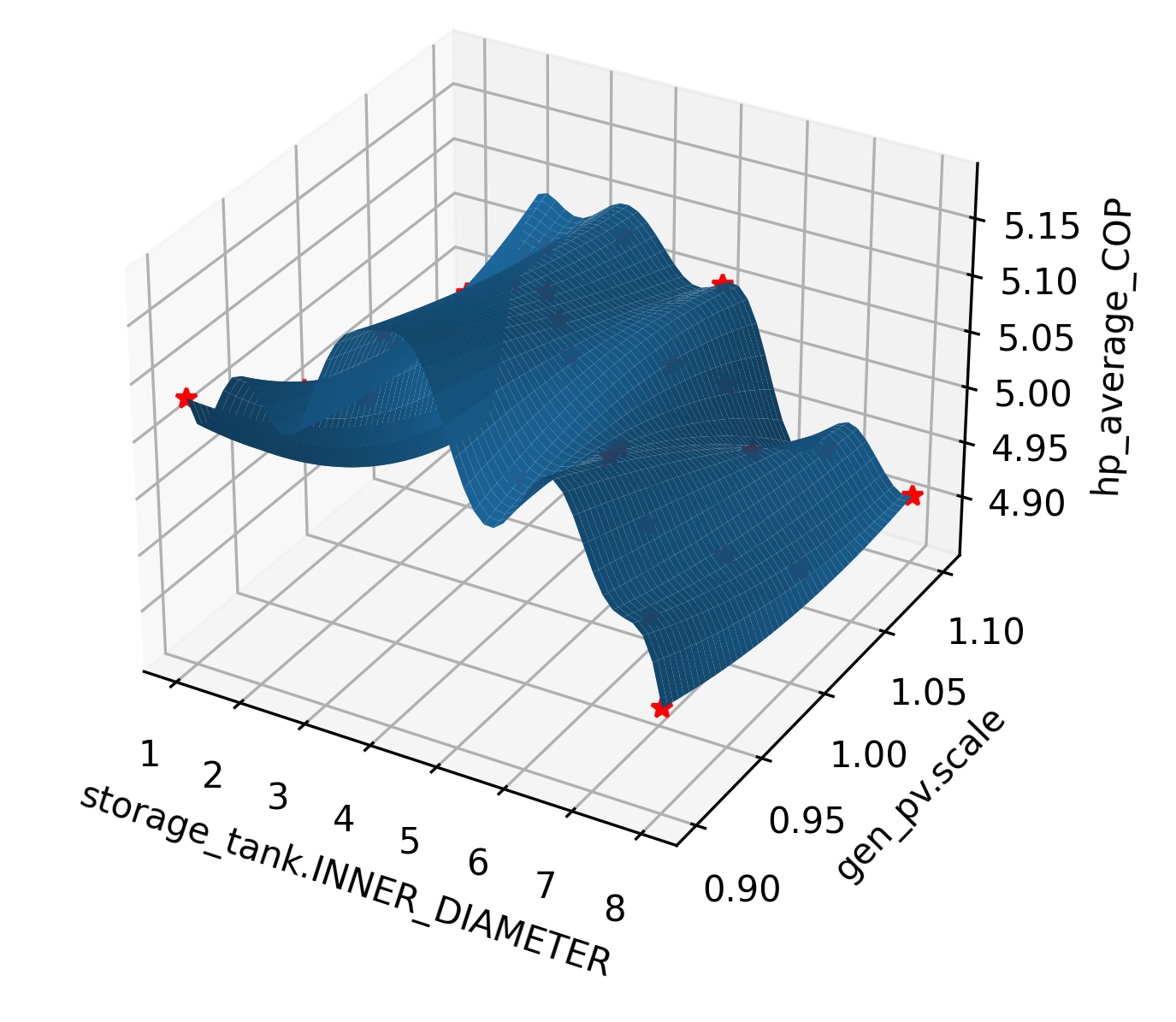}
        \caption{Average COP of heat pump.}
    \label{fig:meb_scaling_pv_averageCOP}
    \end{subfigure}
    \newline
    \begin{subfigure}{.23\textwidth}
        \centering
        \includegraphics[width=1\textwidth]{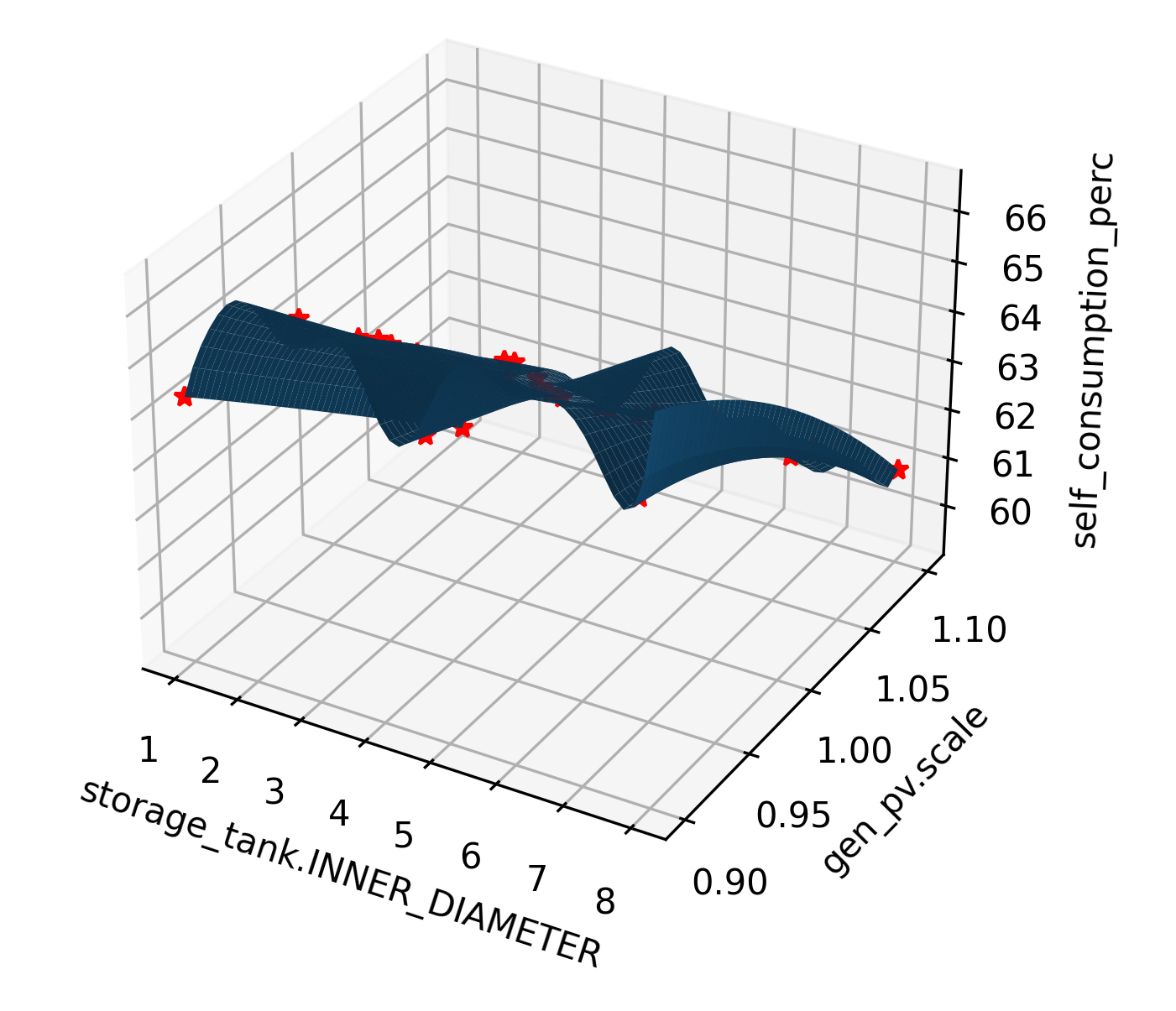}
        \caption{Self-consumption in percent.}
        \label{fig:meb_scaling_pv_selfConsumption}
    \end{subfigure}
    \begin{subfigure}{.24\textwidth}
        \centering
        \includegraphics[width=1\textwidth]{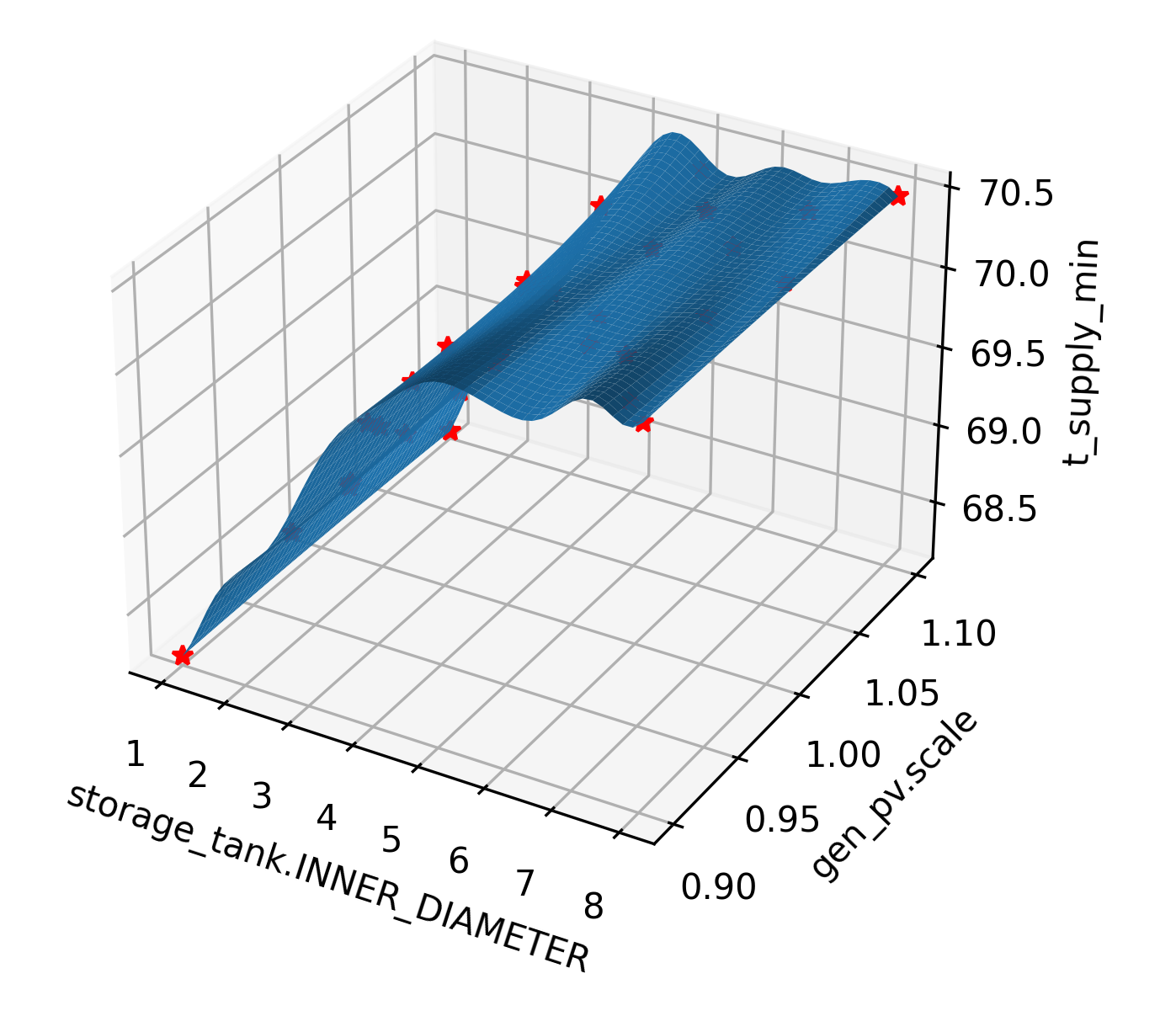}
        \caption{Minimum supply temperature.}
        \label{fig:meb_scaling_pv_Tsupplymin}
    \end{subfigure}
    \caption{Impact of the inner diameter scaling of the Hot Water Tank (HWT) and size scaling of the PV systems to four target metrics.}
    \label{fig:Impact of variations in the inner diameter of the HWT and scaling of the PV systems to four target metrics}
\end{figure}
Figure \ref{fig:Impact of variations in the inner diameter of the HWT and scaling of the PV systems to four target metrics} displays the analysis of the inner diameter of the HWT in combination with the scaling of PV systems. The relative impact of PV scaling on both voltage and self-consumption appears to significantly surpass the impact of HWT size scaling, as demonstrated by the trends seen in Figure \ref{fig:Impact of variations in the inner diameter of the HWT to four target metrics.}. In this combined analysis, no consistent effect of the tank size is apparent. Conversely, for the average COP and the critical node temperature, the scaling of the PV system seems to have a less significant effect, and the shapes are similar to those observed in Figure \ref{fig:Impact of variations in the inner diameter of the HWT to four target metrics.}. The meta-models exhibit wavy effects, particularly evident in the average COP in Figure \ref{fig:meb_scaling_pv_averageCOP}. While these effects have not been fully explained, they are recognized as modeling artefacts. A hypothesis to explore further is the potential impact of the chosen order of the meta-model, combined with discretization effects of the simulation model utilized in this study.
\begin{figure}[t]
    \centering
    \begin{subfigure}{.24\textwidth}
        \centering
        \includegraphics[width=1\textwidth]{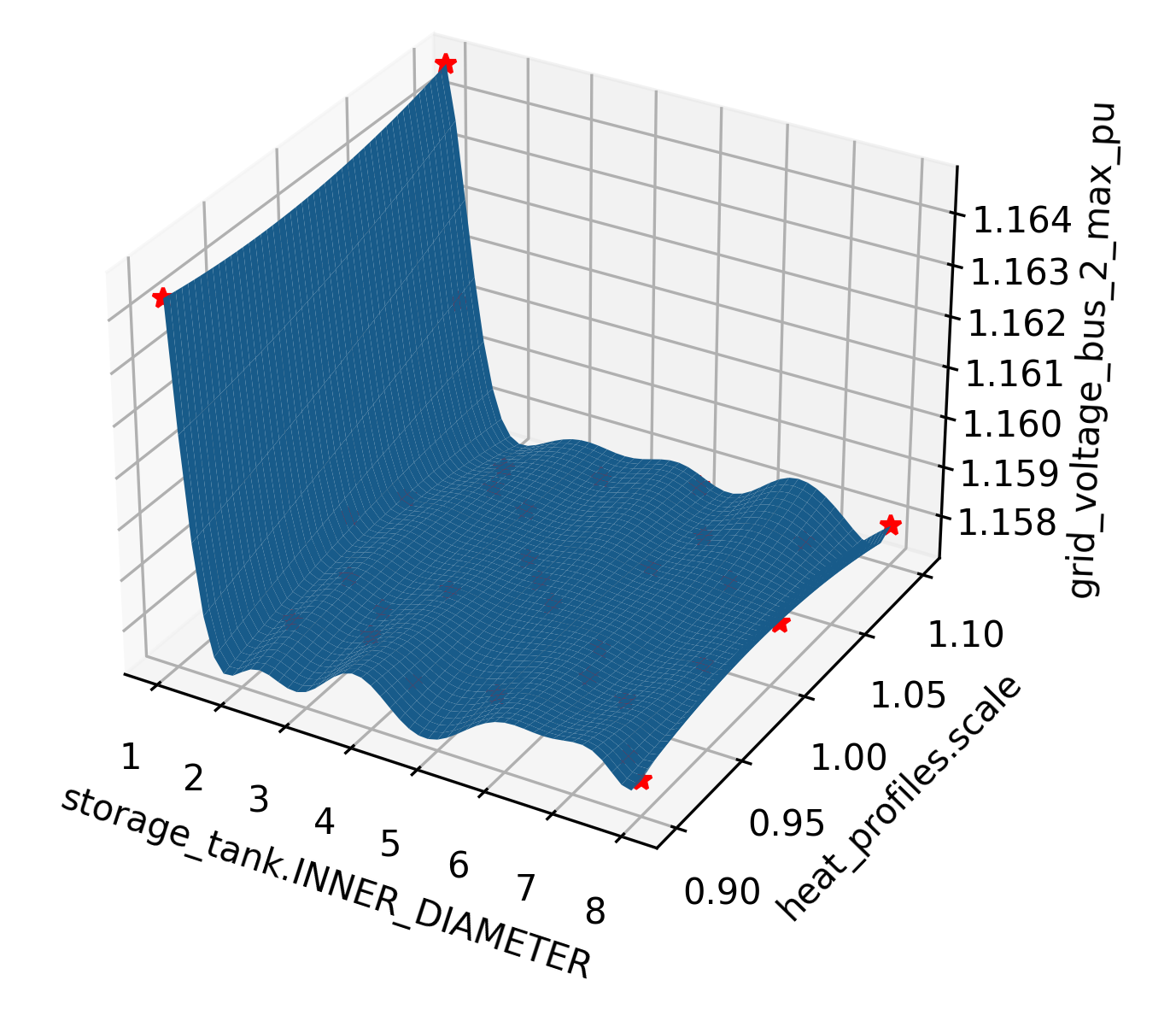}
        \caption{Maximum voltage at bus 2.}
        \label{fig:HWT_Heat_Scaling_Maximum voltage at bus 2}
    \end{subfigure}
    \begin{subfigure}{.24\textwidth}
        \centering
        \includegraphics[width=1\textwidth]{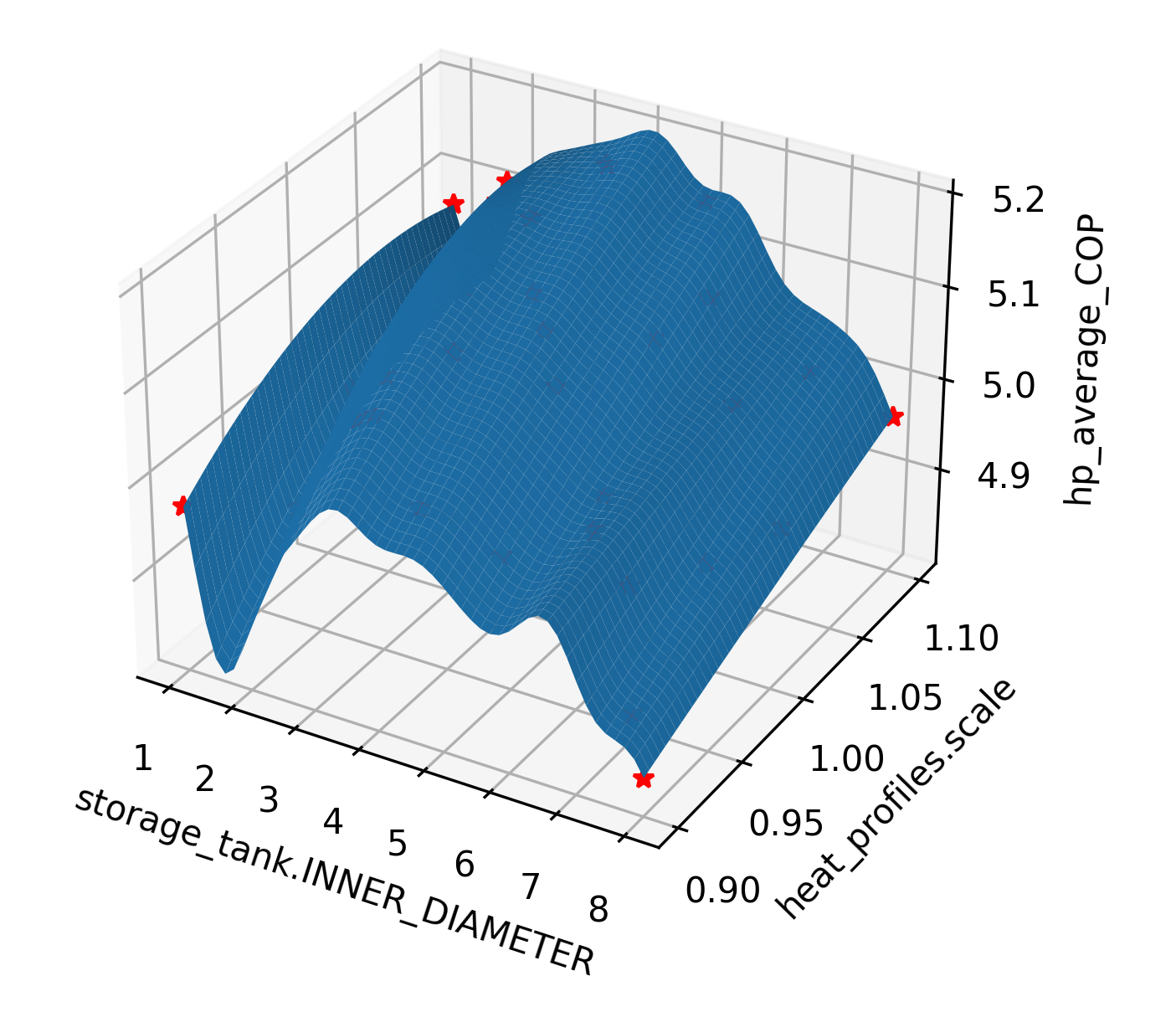}
        \caption{Average COP of heat pump.}
    \label{fig:HWT_Heat_Scaling_COP}
    \end{subfigure}
    \newline
    \begin{subfigure}{.23\textwidth}
        \centering
        \includegraphics[width=1\textwidth]{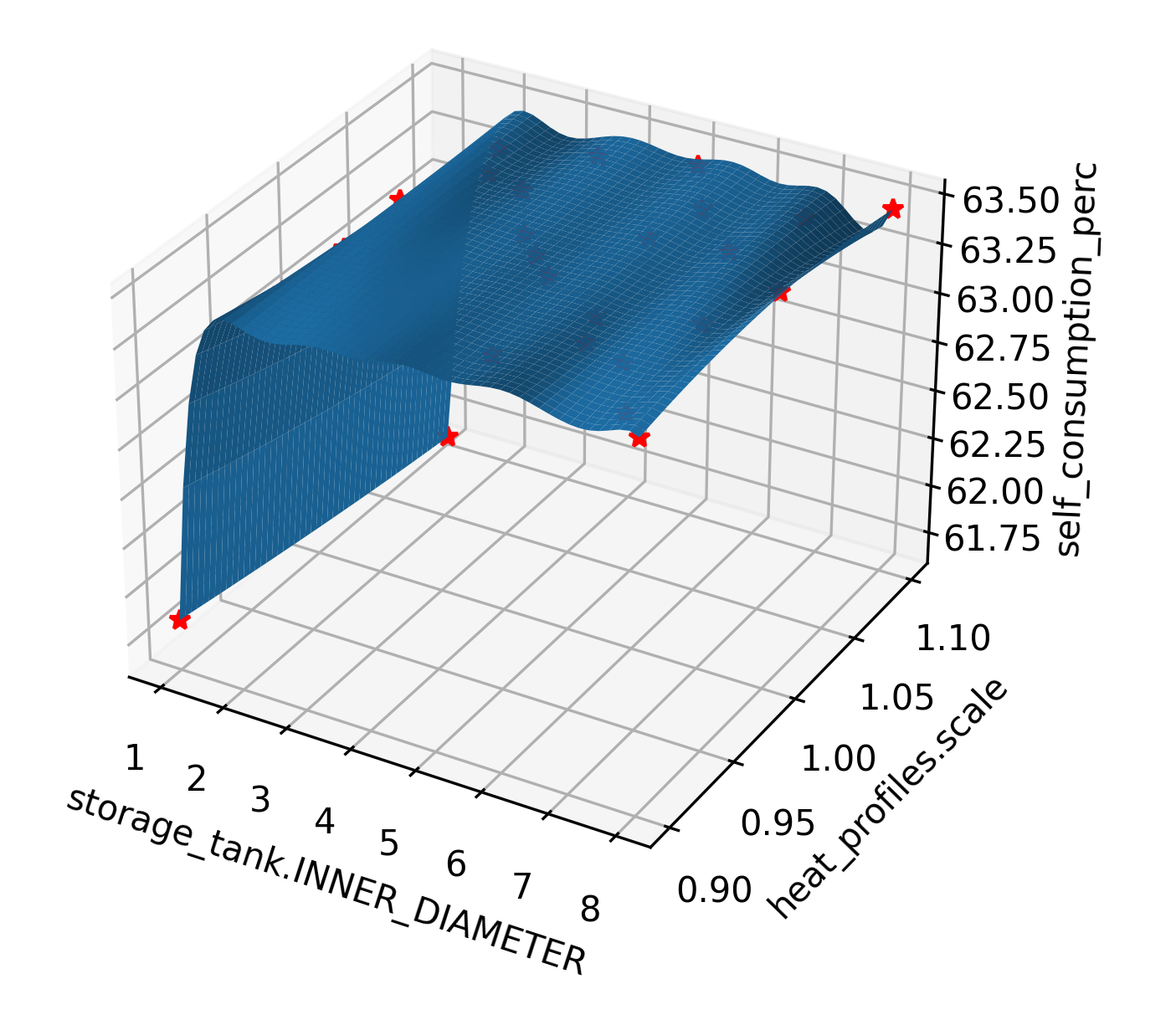}
        \caption{Self-consumption in percent.}
        \label{fig:HWT_Heat_Scaling_Self_Consumption}
    \end{subfigure}
    \begin{subfigure}{.24\textwidth}
        \centering

        \includegraphics[width=1\textwidth]{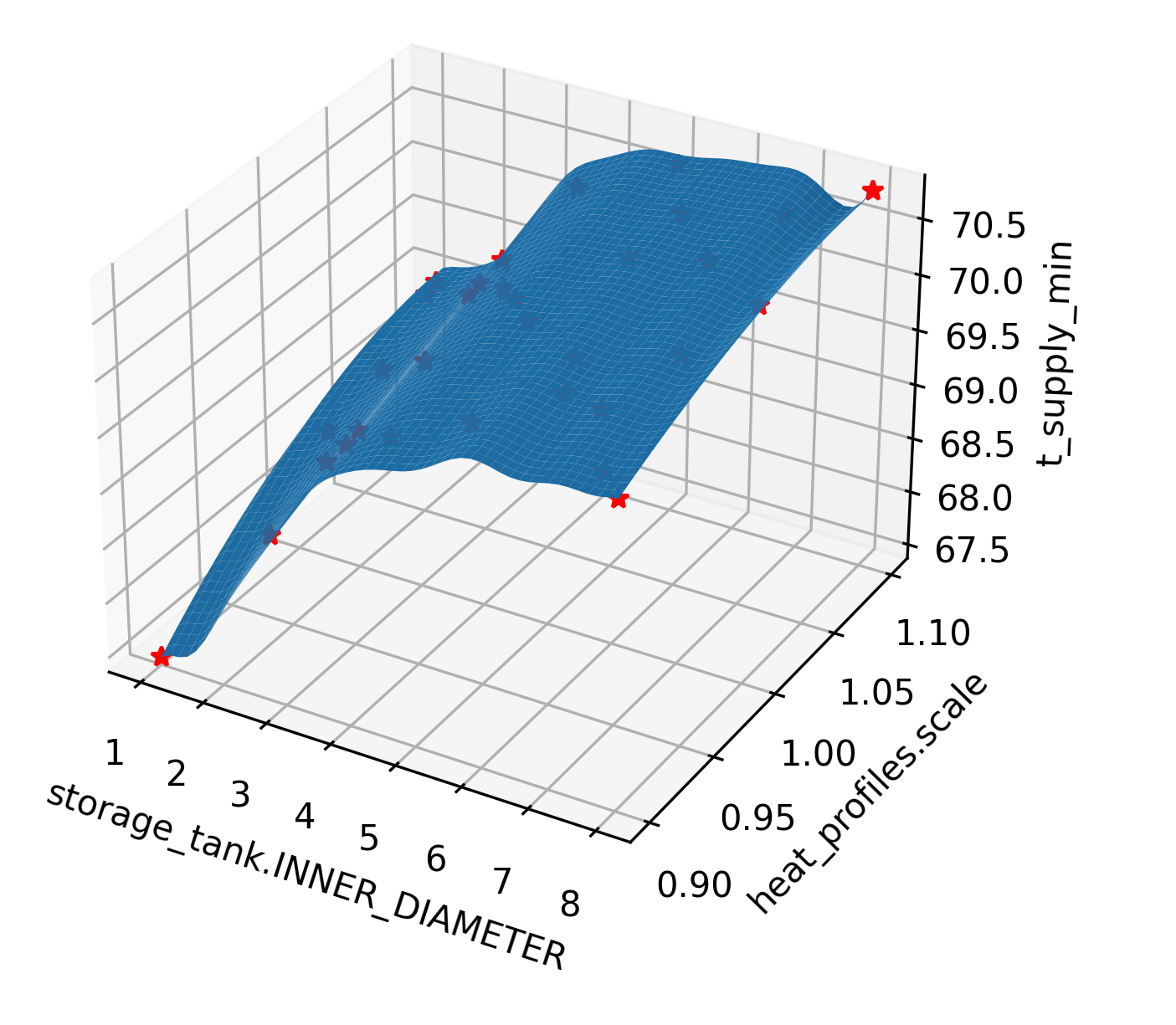}
        \caption{Minimum supply temperature.}
        \label{fig:HWT_Heat_Minimum supply temperature}
    \end{subfigure}
    \caption{Impact of the inner diameter scaling of the HWT and the heat profile scaling to four target metrics.}
    \label{fig:Impact of scaling with the inner diameter of the HWT and the heat profile systems
to four target metrics.}
\end{figure}

Figure \ref{fig:Impact of scaling with the inner diameter of the HWT and the heat profile systems to four target metrics.} presents the simulation results with varied factors: the inner diameter of the HWT and the heat profile scaling. It is evident that the heat profile scaling has a lesser impact on both voltage and self-consumption, consistent with the trends observed in the one-factor analysis shown in Figure \ref{fig:Impact of variations in the inner diameter of the HWT to four target metrics.}. The tendency from the sensitivity analysis in the previous section is reaffirmed here, with Figure \ref{fig:Sobol indices for critical node temperature of heat} illustrating that the heat profile scaling has a significantly greater impact on the critical node temperature compared to the PV scaling. However, the impact of the inner diameter remains stronger, although there is some observable influence of the heat profile scaling. After considering the initial design question of how the tank size impacts a) mitigating voltage problems and b) maximizing self-consumption of renewable energy in the area, it can be seen that the impact of the tank size on the specified target metrics is limited to specific intervals. Further investigation is required to determine if modifying the current control strategy would yield different outcomes.

\section{Conclusion and Outlook}
This paper addresses multi-energy system sizing, considering "domain expansion" and "scaling." By merging these aspects, it offers insights into interactions and challenges tied to expanding and sizing such systems for sustainable energy planning. The study presents two methods for the examined cases, highlighting heat-electricity domain interactions and HWT sizing implications. The first involves OAT sensitivity analysis, dissecting individual factors, followed by a GSA to further investigate effects of the chosen factors. The second uses meta-modeling, creating higher-level models to capture intricate interactions comprehensively. 

Moreover, the simulation results guide the identification of physical testing needs for physical laboratory testing and the design's field roll-out. Initially, the analysis enhances individual component models, followed by validating predictions from the combined model.

\end{document}